%% file: paper.tex
\documentclass[11pt,psfig]{article}  
\usepackage{a4wide}
\usepackage{graphicx}
\usepackage{subfigure}
\usepackage[centertags]{amsmath}
\usepackage{amssymb}

\newcommand{\Z}{{\mathbb Z}}
\newcommand{\R}{\mathcal R}
\newcommand{\OR}{\Omega {\mathcal R}}

\newcommand{\ku}{\mathcal{K}}
\newcommand{\au}{\mathcal{A}}
\newcommand{\msu}{\mathcal{M}}
\newcommand{\lu}{\mathcal{L}}
\newcommand{\kt}{\Tilde{\mathcal{K}}}
\newcommand{\at}{\Tilde{\mathcal{A}}}
\newcommand{\mt}{\Tilde{\mathcal{M}}}
\newcommand{\lt}{\Tilde{\mathcal{L}}}

\newcommand{\AN}{{\mathcal A}}
\newcommand{\MS}{{\mathcal M}}

\newcommand{\LAT}{{\mathcal L}}
\newcommand{\KB}{{\mathcal K}}
\newcommand{\SC}{\scriptstyle}
\newcommand{\Anti}{\text{\bf A}}
\newcommand{\Sym}{\text{\bf S}}

\newcommand{\F}{\text{\bf F}}

\newcommand{\3}{{\bf 3}}
\newcommand{\2}{{\bf 2}}

\newcommand{\barre}[1]{%
        \setbox1=\hbox{$#1$} \dimen2=\ht1 \dimen3=\dp1 \dimen4=\wd1  
        \setbox2=\hbox{\sl /}  
        \dimen1=\wd1 \advance\dimen1 by -\wd2 \divide\dimen1 by 2  
        \advance\dimen1 by \wd2 \advance\dimen1 by 0.4pt  
        \setbox3=\hbox to \wd1{\hss \box1 \kern -\dimen1 \box2\hss}  
        \ht3=\dimen2 \dp3=\dimen3 \wd3=\dimen4  
        \box3  
        }  
\begin{document}

\pagestyle{empty}  
\begin{flushright}  
                     hep-th/0105208 
\end{flushright}  
\vskip 2cm    

\begin{center}  
{\huge Orientifolds with branes at angles}  
\vspace*{5mm} \vspace*{1cm}   
\end{center}  
\vspace*{5mm} \noindent  
\vskip 0.5cm

\centerline{{\bf Stefan F\"orste}, {\bf Gabriele Honecker}, {\bf Ralph
 Schreyer}}

\vskip 1cm
\centerline{\em Physikalisches Institut, Universit\"at Bonn}
\centerline{\em Nussallee 12, D-53115 Bonn, Germany}
\vskip2cm
  
\centerline{\bf Abstract}  
We present supersymmetry breaking four dimensional orientifolds of
type IIA strings. The compact space is a torus times a four dimensional
orbifold. The orientifold group reflects one direction in each
torus. RR tadpoles are cancelled by D6-branes intersecting at angles in
the 
torus and in the orbifold. The angles are chosen such that
supersymmetry is broken. The resulting four dimensional theories
contain chiral fermions. The tadpole cancellation conditions imply
that there are no non-abelian gauge anomalies. The models also contain
 anomaly-free  $U(1)$ factors.

\vskip .3cm

  
\newpage  

\setcounter{page}{1} \pagestyle{plain}  

\section{Introduction} \label{intro}

One of the very important problems of string theory is the existence
of a huge number of different vacua. As long as no dynamical
mechanisms to select only a few (or even a unique) vacuum are
known, one reasonable criterion to search for the correct vacuum is that
in the low energy limit it should look similar to observations. This
means that the resulting effective field theory should be the standard
model or something ``slightly beyond'' the standard model. For a long
time this criterion selected out especially Calabi-Yau
compactifications of the ten dimensional heterotic string. These
result in $N=1$ supersymmetric gauge theories coupled to gravity,
which (as a category of theories) is fairly close to the standard
model. After clarifying the importance of
D-branes\cite{Polchinski:1995mt} and O-planes, the heterotic string lost its
inimitability as a provider of phenomenologically interesting models. 
Now, gauge groups containing the standard model groups can appear due
to open strings ending on D-branes. The appropriate tools to obtain
lower dimensional (e.g.\ four dimensional) effective theories are
orientifold
constructions[\ref{refDbranesa}--\ref{refDbranese}].   

The strategy in constructing orientifold compactifications of type II
strings is as follows. One starts with the compact space being an
orbifold which is often taken to correspond to a limit in the space of
Calabi-Yau manifolds. This breaks the amount of supersymmetry to twice
the minimal amount in the low energy effective theory, and thus is not
only ``slightly'' beyond the amount of supersymmetry in the standard
model. Supersymmetry can be broken by another half when in addition
worldsheet parity (possibly in combination with a discrete target
space mapping) is gauged. Sets of points which are invariant under a
group element containing the worldsheet parity inversion are called
O-planes. These O-planes carry RR charges and typically possess only
compact transverse directions. In order not to contradict RR charge
conservation D-branes need to be added such that the net charge is
zero. (This condition appears as the well known tadpole cancellation
condition.) As long as the D-branes are parallel to the O-planes they
do not break any further supersymmetry. Those orientifold
compactifications give the minimal (apart from zero) amount of
supersymmetry in the low energy effective action. Such models in various
dimensions with and without chiral fermions, with and without D-branes
at angles have been constructed. We give only a non comprehensive list
of references here[\ref{oria}--\ref{orie}].  

In orientifold compactifications it may make sense to push the
supersymmetry breaking one step further, to completely broken
supersymmetry. The reason is that supersymmetry may not be necessary
to explain the hierarchy between the Planck and the electro weak
scale. In brane setups one can (sometimes) keep the fundamental scale
(the string scale) at the weak scale and obtain the Planck scale by
large compact extra dimensions\cite{Antoniadis:1998ig}.

There are essentially two ways to break supersymmetry completely by
D-branes. Instead of satisfying the tadpole cancellation conditions by
adding D-branes only one can also include anti-D-branes (with opposite
RR charges). Strings stretching between anti-D-branes and D-branes can
have tachyonic excitations depending on the distance between the
branes. Under certain circumstances the vacua are then unstable and
undergo phase
transitions[\ref{phasetransa}--\ref{phasetranse}].
Models of this kind have been presented e.g.\
in[\ref{ref_anti-branesa}--\ref{ref_anti-branese}].

In this paper we will be interested in another way of supersymmetry
breaking, namely via branes at angles. D-branes intersecting at angles
generically break supersymmetry completely and only for very special
values of the angles supersymmetry is partly unbroken. Such scenarios
are e.g.\ discussed
in[\ref{ref_branes_at_anglesa}--\ref{ref_branes_at_anglese}]. We will mainly focus on the
constructions in\cite{Blumenhagen:2000wh} and\cite{
  Aldazabal:2000dg}. Indeed, we present basically a combination of
these two models. Let us briefly describe the setup of
\cite{Blumenhagen:2000wh} (we are interested in the four dimensional
model they present). The compact space is a product of three tori
$T^2$. The symmetry which is gauged is worldsheet parity inversion
multiplied with the $\mathbb{ Z}_2$ reflection of three compact
dimensions, one in each torus. This is a symmetry of type IIA
strings. Supersymmetry is completely broken by adding D6-branes which
intersect at angles in each of the three tori. (The D6-branes have one
Neumann and one Dirichlet direction in each torus.) From strings
stretching between D-branes intersecting at angles in all the three
tori chiral fermions are obtained. 
As argued in\cite{Blumenhagen:2000wh,Aldazabal:2000dg} this setup has the disadvantage
that an explanation of the hierarchy by having large extra dimensions
is not possible in this model. The reason is that there is no
``overall'' transverse dimension which could be taken large. We will
comment on this criticism below. But first we want to sketch the
alternative model presented in\cite{Aldazabal:2000dg}. Here, the
compact space is chosen to be a $T^2$ times a four dimensional
orbifold. The model contains D4-branes which are extended along the
non-compact dimensions and in one direction in the $T^2$, where they
intersect at angles. They are located at orbifold singularities in the
remaining four compact directions. Again, chiral fermions arise from
the sector where a string stretches between two D4-branes intersecting
at angles. The hierarchy between the weak and the Planck scale can now
be translated into a large orbifold volume.

In the present paper we combine these two constructions, i.e.\ we
perform a further orbifold gauging of the model
in\cite{Blumenhagen:2000wh}, or (equivalently) an orientifold gauging
of the model in\cite{Aldazabal:2000dg}. The resulting low energy
effective theories are non-supersymmetric and contain chiral
fermions. However, like in the model of\cite{Blumenhagen:2000wh} there
is no overall transverse dimension whose large volume could explain
the hierarchy. But there may be a way out of this problem. The models
are non-supersymmetric and free of RR tadpoles. This necessarily
implies that they possess NSNS tadpoles, or roughly speaking that the
net tension in the transverse space differs from zero. Taking into
account the Fischler-Susskind mechanism\cite{Fischler:1986tb} this
non-vanishing net tension back-reacts on the excitations of the closed
type IIA string, in particular on the metric. Such a back-reaction can
result in a cosmological constant in the effective four dimensional
theory but may also curve the geometry of the compact space. Stringy
setups where this happens have been studied
in\cite{Dudas:2000ff,Blumenhagen:2001dc}. For the kind of setups we
discuss here the effect of the back-reaction has not been taken into
account. This is perhaps a rather complicated
computation. Qualitatively one would expect that the compact space
curves. This affects the size of the four dimensional Planck mass, and
the four dimensional gauge couplings differently. Such an effect may
explain the hierarchy even if one starts without choosing some of the
compact dimensions to be large, for a simple model where this happens
see\cite{Randall:1999ee}. We are not claiming that this will happen in
the models we discuss. However, it may be a way to obtain the scales
in a natural way, about which we just miss sufficiently detailed
knowledge.   

The present paper is organized as follows. In the next section we line
out the general setup. Section 3 provides the details of the
construction. These are the possible torus lattices, the tadpole
cancellation conditions and the massless spectra. In section 4 we
discuss the differences due to the different orbifold projections and
work out some examples explicitly. Section 5 contains a summary of our
results and our conclusions. In two appendices we collect some
formul\ae\ which are used in the loop and tree channel computations of
the tadpole cancellation conditions. A third appendix gives the
massless closed string spectra and chiral fermions due to open strings
in general and for a particular example. 

  
\section{General Setup}
\label{setup}

\begin{figure}[ht]
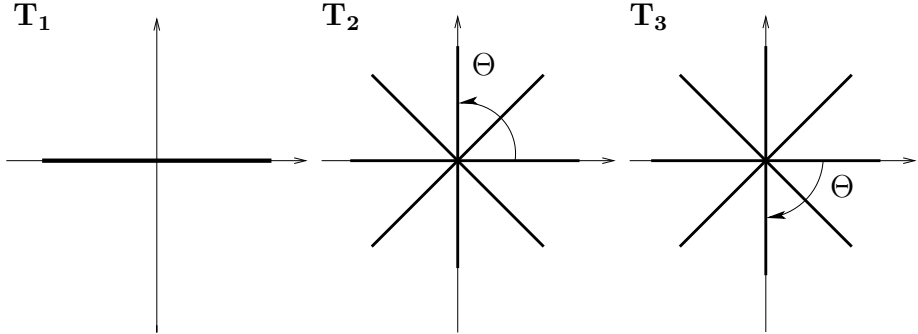
  
\begin{center}
\input o_planes_4.pstex_t
\end{center}
\caption{Orientifold fixed planes for $\Z_4$}
\label{o_planes_4}
\end{figure}

In this article we construct $\OR$ orientifolds of type IIA string
theory compactified to four dimensions on $T^2 \times T^4 / \Z_N$ with
D6-branes at angles. In the following we explain some general features
of these models and introduce some notation. For further details we
refer to~\cite{Forste:2001hx}. The four non-compact dimensions are 
labeled by $x^\mu$, $\mu = 0, \ldots ,3$. The six compact directions
we describe by three complex coordinates,
\begin{equation}
  z^{1}=x^4 + ix^5 \, ,\, z^2 =x^6+ix^7 \, ,\, z^3 = x^8+ix^9 .
\end{equation}
corresponding to three two-tori $T_{1,2,3}$. $\Omega$  
reverses worldsheet parity, ${\cal R}$ reflects the imaginary
parts of the $z^i$,
\begin{equation}
  \R: z^i \rightarrow \bar{z}^i ,
\end{equation}   
and the $\Z_N$ rotation $\Theta$ acts as
\begin{equation} 
  \label{orbifoldaction}
  \Theta :\; z^i \rightarrow e^{2\pi iv_i}z^i ,
\end{equation}
but on the tori $T_{2,3}$ only, i.e. $v_1 =0$.
The set of points invariant under $\OR\Theta^k$ $(k=0, \ldots, N-1)$
constitutes the orientifold fixed planes (O6-planes) which fill the non-compact
dimensions and whose locations in the compact directions are shown in
figure~\ref{o_planes_4}. 
The RR-charges of these O6-planes have to be cancelled by adding an
appropriate arrangement of D6-branes. The most symmetric choice is to
place the D6-branes on top of the O6-planes. This leads to a non-chiral
and supersymmetric theory, in principle given by a simple torus
compactification of the models discussed in~\cite{Blumenhagen:2000md}.
The requirement of balancing the RR-charges of O6-planes and
D6-branes yields the tadpole cancellation
conditions~\cite{Gimon:1996rq} which constitute the basic consistency
conditions for orientifold constructions. 

The tadpole cancellation
conditions can be obtained in two different ways. One possibility is
to extract the relevant UV-divergent parts of the loop channel
diagrams which are given by the trace over the closed string NSNS
states with ${\text\bf P}\OR(-)^F$ insertion, over the open string R
states with $-{\text\bf P}\OR$ and over the open string NS states with
${\text\bf P}(-)^F$ insertion in the case of the Klein bottle, the
M\"obius strip and the annulus, respectively. ${\text\bf P}$ is the 
projector on the states invariant under the orbifold group $\Z_N$ and
$(-)^F$ is the worldsheet fermion number. For further details
again consider~\cite{Forste:2001hx}. The second possibility is to
construct the boundary states $|B\rangle$ for the D6-branes and the
crosscap states $|C\rangle$ for the O6-planes~\cite{Polchinski:1988tu}
and to calculate
directly the exchange of RR closed strings in the tree channel by the
overlaps $\langle C|\ldots|C\rangle$, $\langle C|\ldots|B\rangle$ and
$\langle B|\ldots|B\rangle$ in the case of the Klein
bottle, the M\"obius strip and the annulus, respectively. For further
details, see appendix~\ref{app_tree}.

The requirement that the amplitudes obtained in the two different ways
explained above have to match after a modular transformation from the
loop to the tree channel (or vice versa) is called worldsheet duality
and further restricts the possible solutions. E.g.\ the Klein bottle
amplitude restricts the possible choices of orbifold lattices, the
M\"obius strip yields conditions for the re\-pre\-sentation of the
orientifold group on the Chan Paton matrices and from the
annulus/cylinder one obtains the twisted sector tadpole cancellation
conditions, since the identity $\OR\Theta = \Theta^{-1}\OR$ implies
that only untwisted RR states may propagate in the tree channel.

\section{Details of the construction}
\label{details}

\begin{figure}[ht]
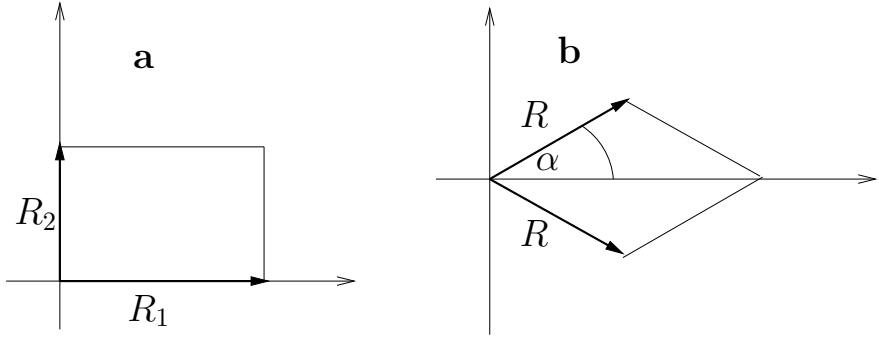
  
\begin{center}
\input torus_1.pstex_t
\end{center}
\caption{Possible torus lattices on $T_1$.}
\label{torus_1}
\end{figure}

\begin{figure}[ht]
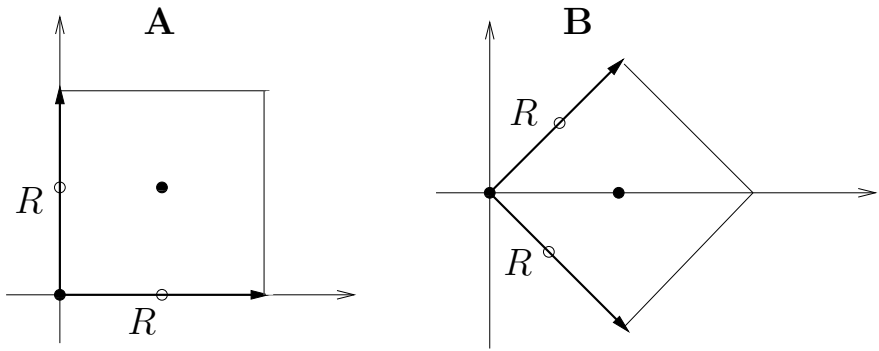
  
\begin{center}
\input torus_23.pstex_t
\end{center}
\caption{Possible torus lattices on $T_{2,3}$ for $\Z_4$. The black
  circles are the fixed points under $\Theta^{1,3}$ and the white
  circles are the additional fixed points under $\Theta^2$.}
\label{torus_23}
\end{figure}

\begin{figure}[ht]
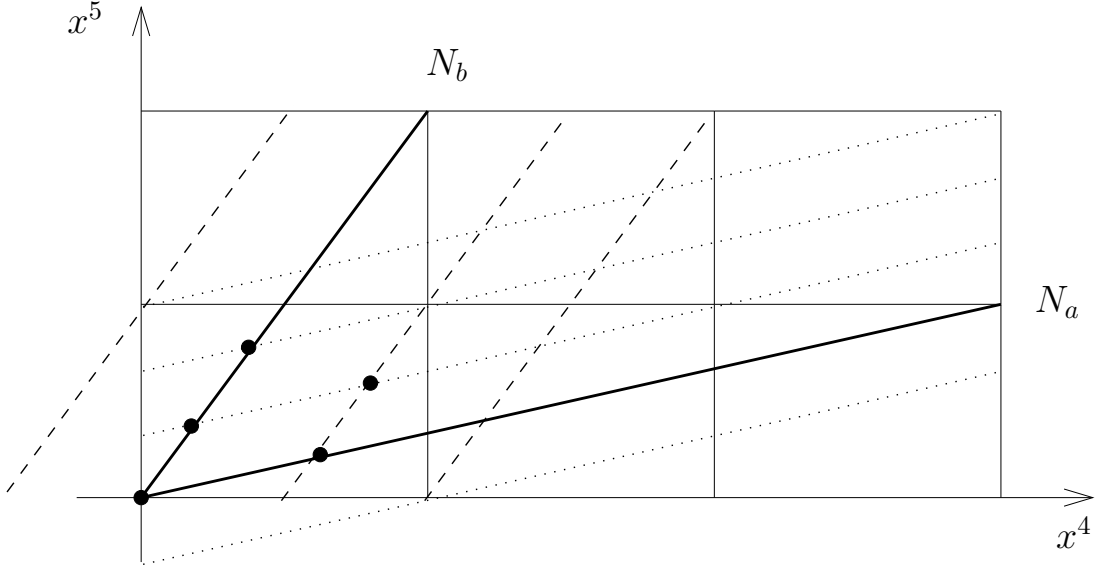
  
\begin{center}
\input angles.pstex_t
\end{center}
\caption{One stack of $N_a$ D6$_a$-branes with wrapping number $(n_a,m_a)
  = (3,1)$ intersecting another stack of $N_b$ D6$_b$-branes with wrapping
  number $(n_b,m_b) = (1,2)$ on $T_1$. The black circles denote the
  multiple intersection points on the torus.}
\label{angles}
\end{figure}

Before presenting any calculations, we describe some general properties.
The tori have to be invariant under the orientifold group, i.e. the
torus $T_1$ has to be invariant under $\R$ whereas the tori $T_{2,3}$
have to be $\Z_N$ invariant in addition. The action of
$\Z_N$ is defined via the shift $\vec{v} = (0,1/N,-1/N)$. The possible
lattices are shown in the figures \ref{torus_1} and \ref{torus_23} for
the example of $\Z_4$. In the case of $\Z_2$, the tori $T_{2,3}$ can be
chosen to be of type {\bf a} (or {\bf b}), too.
%
%
%
We consider $K$ stacks of branes intersecting at arbitrary angles on
$T_1$. Each stack consists of $N_a$ $(a=1,\ldots,K)$ D6$_a$-branes on top
of each other. In figure~\ref{angles}
an example is shown. On $T_{2,3}$ we choose the D6$_a$-branes to be arranged
symmetrically (recall figure~\ref{o_planes_4}).  
%
%
The relative angles of the D6$_a$-branes with respect
to the $x^4$ axis can be expressed in terms of the wrapping numbers
$(n_a,m_a)$, i.e. the number of times the D6$_a$-branes wrap around the
two fundamental cycles of $T_1$.
We only consider models where the angles can be expressed in
finite $(n_a,m_a)$, because otherwise the D6$_a$-brane would fill the
$T_1$. Since $\R$ is a symmetry of the theory, the branes we have to
consider actually are superpositions of D6$_a$-branes with their mirror branes
with respect to the $x^4$ axis. Therefore we may choose
$m_a \ge 0$. Furthermore, the choice $n_a > 0$ reflects the fact that
we do not consider $\overline{\text D6}_a$-branes in this paper. As one can
see in figure~\ref{angles}, in general the branes intersect plurally
on $T_1$. The number of intersections of two stacks of branes D6$_a$
and D6$_b$ can be expressed in terms of the wrapping numbers
\begin{equation}
  \label{intersection_number}
  I_{ab} = n_a m_b - m_a n_b ,
\end{equation}
as one can check for the example in figure~\ref{angles}. A negative
wrapping number makes no sense geometrically, but one can get rid of
the negative sign by choosing the conjugate representation for the
corresponding particles.


\subsection{Amplitudes and tadpole cancellation}
\label{amplitudes}

We start with the Klein bottle loop channel amplitude. The compact
momenta consist of Kaluza Klein contributions
\begin{equation}
  \label{kk_kb}
  P = r_1 \vec{e}_1^* + r_2 \vec{e}_2^*
\end{equation}
and contributions from windings
\begin{equation}
  \label{windings_kb}
  \alpha' W = s_1 \vec{e}_1 + s_2 \vec{e}_2 ,
\end{equation}
where the $r_i$, $s_i$ are integers and the $\vec{e}_i$
($\vec{e}_i^*$) are the basis vectors of the (dual) torus
lattice. Requiring invariance of $P$ and $W$ under the orientifold
group and using $\Omega P \Omega^{-1} = P$ and $\Omega W \Omega^{-1} =
-W$ one obtains the lattice contributions to the amplitude 
$\LAT^{R_1,R_2}[1,1]$ for the {\bf a} type $T_1$ and
$\LAT^{R,R}[1/\cos^2\alpha,4\sin^2\alpha]$ for the {\bf b} type
$T_1$. On $T_{2,3}$ one has to set $R_1=R_2=R$ for {\bf A} and
$\alpha=\pi/4$ for {\bf B} in the case of $\Z_4$, for example (see
appendix~\ref{app_loop_lattice} for the notation). Of course, in
general the radii 
of the different tori are not the same, so there should be further
indices for the radii, but we omit them to keep the notation
simple. The different lattice 
contributions are summarized in table~\ref{lattice_table}. 
Lattice contributions only show up in the untwisted sector.
One could think that all possible combinations {\bf aAA}, {\bf bAA}
etc. are allowed, but here worldsheet duality comes into the game,
see~\cite{Blumenhagen:2000md,Forste:2001hx} and appendix~\ref{app_tree}. In the
case of $\Z_4$ (and $\Z_6$), an insertion of $\Theta$ in the trace
interchanges contributions from {\bf A} and {\bf B}. But since, as
stated above, in the 
tree channel only untwisted RR states propagate, each contribution in
the loop channel has to be invariant under interchange of {\bf A} and {\bf
  B}. Thus in the cases of $\Z_{4,6}$ only the models {\bf
  aAB} and {\bf bAB} (which are equivalent to {\bf aBA} and {\bf bBA},
respectively), whereas for $\Z_{2,3}$ all combinations are
allowed. 

The calculation of the NSNS part with $(-)^F$ insertion of the one
loop amplitude yields
\begin{equation}
  \label{kb_amplitude_loop}
  \ku =c \int_0^{\infty}\frac{dt}{t^3} 
  \lu^{\ku}_1\left(\ku^{(0)}\lu^{\ku}_2\lu^{\ku}_3+\sum_{n=1}^{N-1}
    \chi^{(n)}\ku^{(n)}\right),
\end{equation}
where $\ku^{(n)}$ are the oscillator contributions from the $n$th
twisted sector (see~\cite{Blumenhagen:2000ev,Forste:2001hx} for the
notation), $\chi^{(n)}$ are the 
multiplicities from the orbifold fixed points and $c = V_4/(8\pi^2\alpha')^2$
is the regularized volume factor from the non-compact momentum
integration. Note that the result~(\ref{kb_amplitude_loop}) already
implies the consequences from worldsheet duality, namely the fact that
not all combinations of lattices are allowed, as stated above.

Performing the modular transformation using $t=1/4l$ leads to the tree
channel amplitude
\begin{equation}
  \label{kb_amplitude_tree}
  \renewcommand{\arraystretch}{0.7}
  \kt=c c_1^{\ku}c_2^{\ku}c_3^{\ku}\int_0^{\infty} dl 
  \lt^{\ku}_1
  \frac{\vartheta \Bigl[\!\! \begin{array}{c} \SC 1/2 \\ \SC 0
    \end{array} \!\!\Bigr]^2}{\eta^{6}} 
  \left\{ \lt^{\ku}_2\lt^{\ku}_3
    \frac{\vartheta \Bigl[\!\! \begin{array}{c} \SC 1/2 \\ \SC 0
      \end{array} \!\!\Bigr]^2}{\eta^{6}} 
    -4\sum_{n=1}^{N-1}\sin^2\left(\frac{\pi n}{N}\right)
    \frac{\vartheta \Bigl[\!\! \begin{array}{c} \SC 1/2 \\ \SC n/N
      \end{array} \!\!\Bigr]  
      \vartheta \Bigl[\!\! \begin{array}{c} \SC 1/2 \\ \SC -n/N \end{array}
      \!\!\Bigr]} 
    {\vartheta \Bigl[\!\! \begin{array}{c} \SC 1/2 \\ \SC 1/2+n/N \end{array}
      \!\!\Bigr] 
      \vartheta \Bigl[\!\! \begin{array}{c} \SC 1/2 \\ \SC 1/2-n/N
      \end{array} \!\!\Bigr]} 
  \right\},
\end{equation}
corresponding to the RR exchange in the tree channel. The $c_i^{\ku}$
can be read off from $\lu_i^{\ku} = c_i^{\ku} l \lt_i^{\ku}$ and are
displayed in table~\ref{lattice_table}. The
argument of the $\vartheta$ and $\eta$ functions in
equation~(\ref{kb_amplitude_tree}) is given by $e^{-4\pi l}$.

Calculating the amplitude directly in the tree channel using the
boundary state formalism~\cite{Polchinski:1988tu} and comparing the
result with equation~(\ref{kb_amplitude_tree}) (i.e.\ requiring
worldsheet duality) restricts the possible choice of the tori, as
stated above, and determines the normalization of the crosscap states
to be 
\begin{equation}
  \label{normal_c}
  {\cal N}_C = \sqrt{\frac{c c_1^{\ku} c_2^{\ku} c_3^{\ku}}{2N}},
\end{equation}
see appendix~\ref{app_tree_crosscap} for the details and notation.

Now we consider the annulus amplitude, i.e.\ the one-loop amplitude
from strings starting and ending on some of the D6-branes we have to
add to cancel the tadpoles arising from the orientifold
planes. Remember that we add $N_a$ D6$_a$-branes $(a=1, \ldots, K)$ at
some angles on $T_1$, $\Z_N$ symmetric on $T_{2,3}$. Let us first
discuss the one-loop amplitude from strings starting and ending on the
same stack of D6$_a$-branes. The compact momenta corresponding to the
Kaluza Klein momenta in the case of the Klein bottle are given by the
inverse length of the D6-branes on the corresponding torus. Thus,
e.g.\ for the {\bf a} type $T_1$ one obtains
\begin{equation}
  \label{kk_an}
  P = \frac{r}{\sqrt{(n_a R_1)^2 + (m_a R_2)^2}} \equiv r V^{-1},
\end{equation}
where $r$ is an integer. Strictly speaking, $P$ and $V$ should carry
the index of the wrapping numbers, but we omit this index here and in
the following.
The compact momenta corresponding to the
windings in the case of the Klein bottle are given by the distance of
parallel D6-branes on the corresponding torus, see
figure~\ref{angles}. For the {\bf a} type $T_1$ one obtains
\begin{equation}
  \label{windings_an}
  \alpha' W = s R_1 R_2 V^{-1},
\end{equation}
with $s$ integer. The lattice contributions then appear as
$\lu^{\au}_{\text{\bf a}} = \sum_{r,s}
e^{-2\pi t \alpha' M^2}$ with $M^2 = P^2 + (\alpha' W)^2$ in the
trace. The results for the {\bf b} type $T_1$ are shown in
table~\ref{lattice_table}. The relevant part of
the one loop amplitude from D6$_a$-D6$_a$ strings reads 
\begin{equation}
  \label{an_amplitude_loop_aa}
  \renewcommand{\arraystretch}{0.7}
  \au_{aa} = \frac{c}{4} N_a^2 \int_0^{\infty} \frac{dt}{t^3} 
  \lu_1^{\au}  
  \frac{\vartheta \Bigl[\!\! \begin{array}{c} \SC 0 \\ \SC 1/2
  \end{array}
    \!\!\Bigr]^2}{\eta^{6}}
  \left\{ \lu_2^{\au} \lu_3^{\au}
    \frac{\vartheta \Bigl[\!\! \begin{array}{c} \SC 0 \\ \SC 1/2 \end{array}
      \!\!\Bigr]^2}{\eta^{6}}
    + \sum_{n=1}^{N-1} \chi_{\au}^{(n)}
    \frac{\vartheta \Bigl[\!\! \begin{array}{c} \SC n/N \\ \SC 1/2
      \end{array} \!\!\Bigr]  
      \vartheta \Bigl[\!\! \begin{array}{c} \SC -n/N \\ \SC 1/2 \end{array}
      \!\!\Bigr]} 
    {\vartheta \Bigl[\!\! \begin{array}{c} \SC 1/2+n/N \\ \SC 1/2 \end{array}
      \!\!\Bigr] 
      \vartheta \Bigl[\!\! \begin{array}{c} \SC 1/2-n/N \\ \SC 1/2
      \end{array} \!\!\Bigr]}
  \right\},
\end{equation}
where $\chi_{\au}^{(n)}$ is the intersection number of the D6-branes
on $T_{2,3}$. The
argument of the $\vartheta$ and $\eta$ functions is $e^{-\pi t}$.
Equation~(\ref{an_amplitude_loop_aa}) already contains 
another consequence of worldsheet duality, namely the fact 
that the Chan Paton representations of $\Z_2$ elements
of the orbifold group have to be traceless. This is the so called
``twisted tadpole cancellation condition''.

The modular transformation $t=1/2l$ leads to
\begin{equation}
  \label{an_amplitude_tree_aa}
  \renewcommand{\arraystretch}{0.7}
  \at_{aa} = \frac{c}{2^4} N_a^2 c_1^{\au} c_2^{\au} c_3^{\au} 
  \int_0^{\infty} dl
  \lt^{\au}_1
  \frac{\vartheta \Bigl[\!\! \begin{array}{c} \SC 1/2 \\ \SC 0
    \end{array} \!\!\Bigr]^2}{\eta^{6}} 
  \left\{ \lt^{\au}_2\lt^{\au}_3
    \frac{\vartheta \Bigl[\!\! \begin{array}{c} \SC 1/2 \\ \SC 0
      \end{array} \!\!\Bigr]^2}{\eta^{6}} 
    -4\sum_{n=1}^{N-1} \ldots
  \right\},
\end{equation}
where $-4\sum_{n=1}^{N-1} \ldots$ is just the same as in
equation~(\ref{kb_amplitude_tree}). The
argument of the $\vartheta$ and $\eta$ functions is $e^{-4\pi l}$.
Comparing this with the result from the boundary state formalism, one
obtains the normalization factor for the boundary states
\begin{equation}
  \label{normal_b}
  {\cal N}_B = \sqrt{\frac{c c_1^{\au} c_2^{\au} c_3^{\au}}{2^5N}},
\end{equation}
see appendix~\ref{app_tree_boundary} for the details and notation.

Having found the complete boundary and crosscap states, the
calculation of the remaining amplitudes becomes a straightforward
task. The tree level cylinder amplitude from strings stretched between
the branes D6$_a$ and D6$_b$ intersecting at a relative angle
$\pi\Delta\varphi$ reads
\begin{equation}
  \label{an_amplitude_tree_ab}
  \renewcommand{\arraystretch}{0.7}
  \at_{ab} = \frac{c}{2^3}N_aN_bI_{ab}c_2^{\au}c_3^{\au}
  \int_0^{\infty}dl
  \frac{\vartheta\Bigl[\!\!
      \begin{array}{c}  \SC 1/2 \\
        \SC 0 \end{array}\!\!\Bigr]}{\eta^3}
  \frac{\vartheta\Bigl[\!\!
      \begin{array}{c} \SC 1/2 \\ \SC \Delta\varphi \end{array}\!\!\Bigr]}
  {\vartheta\Bigl[\!\! 
      \begin{array}{c} \SC 1/2 \\
        \SC 1/2+\Delta\varphi \end{array}\!\!\Bigr]}
  \left\{ \lt^{\au}_2\lt^{\au}_3
    \frac{\vartheta \Bigl[\!\! \begin{array}{c} \SC 1/2 \\ \SC 0
      \end{array} \!\!\Bigr]^2}{\eta^{6}} 
    -4\sum_{n=1}^{N-1} \ldots
  \right\},
\end{equation}
where $I_{ab}$ is the intersection number of D6$_a$ and D6$_b$ on
$T_1$ defined in equation~(\ref{intersection_number}).

The contributions from the M\"obius strip to the RR exchange in the
tree channel are obtained by calculating the overlap of the
corresponding boundary and crosscap states. We first discuss the case
of a string starting on a D6-brane aligned with the $x^4$-axis on
$T_1$ and ending on the corresponding O6-plane (and vice versa). The
relevant part of the amplitude is given by
\begin{equation}
  \label{mo_amplitude_tree}
  \renewcommand{\arraystretch}{0.7}
  \mt_{||} = -\frac{c}{2^4} N_a c_1^{\msu} c_2^{\msu} c_3^{\msu} 
  \int_0^{\infty} dl
  \lt^{\msu}_1
  \frac{\vartheta \Bigl[\!\! \begin{array}{c} \SC 1/2 \\ \SC 0
    \end{array} \!\!\Bigr]^2}{\eta^{6}} 
  \left\{ \lt^{\msu}_2\lt^{\msu}_3
    \frac{\vartheta \Bigl[\!\! \begin{array}{c} \SC 1/2 \\ \SC 0
      \end{array} \!\!\Bigr]^2}{\eta^{6}} 
    -4\sum_{n=1}^{N-1} \ldots
  \right\}.
\end{equation}
The argument of the $\vartheta$ and $\eta$ functions is $-e^{-4\pi l}$.
Similarly, we obtain the relevant contribution from a string starting
(or ending) on a D6$_a$-brane at angle $\pi\varphi$ with respect to the
$x^4$-axis on $T_1$
\begin{equation}
  \label{mo_amplitude_tree_a}
  \renewcommand{\arraystretch}{0.7}
  \mt_a = -\frac{c}{2^2}N_aI_{a'a}^{\OR}c_2^{\msu}c_3^{\msu}
  \int_0^{\infty}dl 
  \frac{\vartheta\Bigl[\!\!
      \begin{array}{c}  \SC 1/2 \\
        \SC 0 \end{array}\!\!\Bigr]}{\eta^3}
  \frac{\vartheta\Bigl[\!\!
      \begin{array}{c} \SC 1/2 \\ \SC \varphi \end{array}\!\!\Bigr]}
  {\vartheta\Bigl[\!\! 
      \begin{array}{c} \SC 1/2 \\
        \SC 1/2+\varphi \end{array}\!\!\Bigr]}
  \left\{ \lt^{\msu}_2\lt^{\msu}_3
    \frac{\vartheta \Bigl[\!\! \begin{array}{c} \SC 1/2 \\ \SC 0
      \end{array} \!\!\Bigr]^2}{\eta^{6}} 
    -4\sum_{n=1}^{N-1} \ldots
  \right\},
\end{equation}
where $I_{a'a}^{\OR}$ is the number of $\OR$-invariant intersections
of the brane D6$_a$ with its $\OR$-image D6$_{a'}$.

Putting everything together and taking the limit $l \rightarrow
\infty$, we obtain the tadpole cancellation conditions 
\begin{equation}
  \label{tcc}
  \begin{aligned}
    \Z_2: \quad & \sum_a N_a n_a = \left\{ 
      \begin{array}{cc} 16 & (\text{\bf aaa}) \\ 8 & (\text{\bf aab})
        \\ 4 & (\text{\bf abb}) \end{array} \right. \\
    \Z_3: \quad & \sum_a N_a n_a = 
      \begin{array}{cc} 4 & (\text{\bf aAA},\text{\bf aAB},\text{\bf aBB})
        \end{array} \\
    \Z_4: \quad & \sum_a N_a n_a = \begin{array}{cc} 8 & (\text{\bf aAB})
        \end{array} \\
    \Z_6: \quad & \sum_a N_a n_a = \begin{array}{cc} 4 & (\text{\bf aAB})
        \end{array}, 
  \end{aligned}
\end{equation}
for the {\bf b} type $T_1$ one just has to replace $n_a$ by $n_a +
m_a$. Since we are restricted to the cases $n_a > 0$ and $m_a \geq 0$,
there is not much freedom left for the construction of
models. In section~\ref{Z-2&z-3models} we give some explicit
examples.


\subsection{The spectrum}
\label{spectrum}

Generically, we obtain non-supersymmetric and chiral models which
contain tachyons. The only exception is the trivial case, where $m_a =
0$ for all $a$, where the spectrum is ${\cal N} = 2$ supersymmetric,
nonchiral and free of tachyons.

In the untwisted closed string sector we find the ${\cal N} = 2$
SUGRA multiplet. The twisted closed string sectors and the sectors
from open strings which do not stretch between D6-branes at angles on
$T_1$ contain further ${\cal N} = 2$ SUSY multiplets. What makes the
spectrum generically non-supersymmetric, are the open strings
stretching between branes at angles on $T_1$. In these sectors we
always find tachyons. We find chiral fermions not only in the sectors
where the strings stretch between branes at angles on all three tori,
but also where they stretch between branes at angles on $T_1$ only,
if an additional $\Z_2$ symmetry from the orbifold group is present,
because in 
this case the left- and right-handed states are distinguished by
different parity under the $\Z_2$ transformation. The appearence
of tachyons can easily be seen considering the mass formula for open
strings in the NS sector
\begin{equation}
  \label{mass_ns}
  \alpha' m^2 = \text{osc} + \frac{1}{2} (\Delta\varphi + 2
    \frac{k}{N} - 1),
\end{equation}
where $-1/2 < \Delta\varphi < 1/2$ is the relative angle on $T_1$ and
$k/N$ is 
the relative angle on $T_{2,3}$ in units of $\pi$. ``osc'' is minus
the moding of the applied oscillator, thus e.g.\ for the state
$\psi^1_{\Delta\varphi - 1/2} |0\rangle$ from a string with $k=0$ we
obtain a tachyon with $\alpha' m^2 = - \Delta\varphi/2$. The
resulting gauge groups are
\begin{equation}
  \label{gauge}
  \begin{aligned}
    \Z_{2,4,6}: \;\;\;\; &
    \prod_{m_a \neq 0} U(N_a/2)^4 \prod_{m_a = 0} U(N_a/2)^2 \\
    \Z_{3}: \;\;\;\;\;\;\;\; &
    \prod_{m_a \neq 0} U(N_a) \prod_{m_a = 0} SO(N_a),
  \end{aligned}
\end{equation}
for further details see appendix~\ref{app_spectra} and the explicit examples
in the following section.


\section{The $\Z_3$ and $\Z_2$ case}\label{Z-2&z-3models}

\subsection{Models on $T^2\times T^4/\Z_3$}\label{Z-3models}

The most simple model on $T^2 \times T^4/\Z_N$ is the case $N=3$ where
the branes lie on top of the O-planes on $T_{2,3}$ (similar to 
figure~\ref{o_planes_4}). In the case of $\Z_3$, $\Theta^{1/2}$ is
a symmetry of the lattice and therefore all branes wrapping the same
cycle on $T_1$ are identified under the orbifold
group~\cite{Forste:2001hx,Blumenhagen:2000md}. Generically, a set of
branes with $(n_a,m_a)$ generates a $U(N_a)$ gauge group whereas for
$(n_a,m_a)=(1,0)$, the branes have an additional
$\OR$-symmetry. Imposing the corresponding projection condition
breaks the gauge group down to $SO(N_a)$. 

We can choose the lattice orientations {\bf A, B} independently on
$T_{2,3}$. The tadpole cancellation condition (\ref{tcc}) is not
affected by this choice, but the chiral fermionic spectrum listed in
table~\ref{Z-3spectrum} receives an overall multiplicity $\chi$ from
the number of brane intersections in the fundamental cell. As $\Theta$
and $\Theta^2$ sectors yield the same contribution, each fermion comes
in an even number of copies.

Strings starting and ending on stacks of branes $a,b$ with different
wrappings on $T_1$ transform in the bifundamental of
$U(N_a)\times U(N_b)$. $\OR$ simply maps $ab$-strings to
$b'a'$-strings. In the special case of strings stretching between mirror
branes $aa'$, we have to analyse the intersection points more
closely. On $T_{2,3}$, all intersections are invariant whereas on
$T_1$ there are $2m_a$ points which are mapped onto themselves by the
reflection $\mathcal{R}$ while the remaining $2m_a(n_a-1)$ form
pairs. The latter support fermions in the antisymmetric $({\bf A}_a)$ and
symmetric $({\bf S}_a)$ representation. At $\OR$ invariant brane
intersections, the symmetric part is projected out and we only obtain
fermions transforming in the antisymmetric representation. The generic
chiral fermionic spectrum as displayed in table~\ref{Z-3spectrum} is
free of purely non-abelian gauge anomalies. This is a consequence of
the tadpole cancellation condition~(\ref{tcc}).

\renewcommand{\arraystretch}{1.5}
\begin{table}[ht]
  \begin{center}
    \begin{equation*}
      \begin{array}{|c||c|c|} \hline
        \multicolumn{3}{|c|}{\rule[-3mm]{0mm}{8mm} \text{\bf
         Chiral fermionic spectrum for $\Z_3$}} \\ \hline\hline
        & \text{rep.} & \text{mult.} 
\\ \hline\hline
aa' & (\Anti_a)_L & 4m_a\chi\\
aa' & (\Anti_a)_L+(\Sym_a)_L & 2m_a(n_a-1)\chi\\
ab & (\Bar{\F}_a,\F_b)_L & 2(n_am_b-n_bm_a)\chi\\
ab' & (\F_a,\F_b)_L & 2(n_am_b+n_bm_a)\chi
\\\hline 
      \end{array}
    \end{equation*}
  \end{center}
 \caption{Generic chiral spectrum for $\Z_3$. $\chi=1(3,9)$ is the
         intersection number on $T_{2,3}$ for the lattices ${\bf AA}$
         (${\bf AB}$, ${\bf BB}$).}
\label{Z-3spectrum}
\end{table}

\subsubsection{Example 1, $\Z_3$}\label{example_1-Z_3}

As we restrict our analysis to D6$_a$-branes, the largest feasible
gauge group which respects the tadpole condition (\ref{tcc}) and
yields chiral fermions is $U(3)\times U(1)$. We can split this group
into $SU(3)\times U(1)^2$.
Choosing the wrapping numbers $(n_1,m_1)=(1,1)$, $(n_2,m_2)=(1,2)$,
out of the two $U(1)$'s the linear combination
\begin{equation}
  Q_{non-an.}=Q_1-\frac{3}{2}Q_2
\end{equation}
is non-anomalous. The remaining $U(1)$ should get
a mass of the order of the string scale by a generalized Green-Schwarz
mechanism involving closed string
moduli~\cite{Aldazabal:2000dg,Lalak:1999bk} (and references therein). The resulting spectrum is
displayed in table~\ref{Ex_1spectrum} for the lattice {\bf aAA}. 
\renewcommand{\arraystretch}{1.5}
\begin{table}[ht]
  \begin{center}
    \begin{equation*}
      \begin{array}{|c||c|c|} \hline
        \multicolumn{3}{|c|}{\rule[-3mm]{0mm}{8mm} \text{\bf
         Chiral spectrum, Ex. 1}} \\ \hline\hline
        & SU(3) \times U(1)_{non-an.}& \text{mult.} 
\\ \hline\hline
11' & ({\bf \Bar{3}})_{2} & 4 \\
12  & ({\bf \Bar{3}})_{-5/2} & 2 \\
12'  & ({\bf 3})_{-1/2} &  6
\\\hline 
      \end{array}
    \end{equation*}
  \end{center}
 \caption{Chiral fermionic spectrum for $\Z_3$ with $(n_1,m_1)=(1,1)$,
         $(n_2,m_2)=(1,2)$ and lattice {\bf aAA}. All
         states are left-handed.}
\label{Ex_1spectrum}
\end{table}


\subsection{Models on $T^2\times T^4/\Z_2$}\label{Z-2models}

$\Z_2$ models are the most simple examples for $T^4/\Z_M$ orbifolds
where $M$ is even. In this case, the requirement of obtaining the
complete projector in the tree channel leads to two tadpole
cancellation conditions, namely
\begin{eqnarray}
\sum_a\chi n_aN_a &= &16,\nonumber\\
\text{tr}\gamma_{\Theta}& = &0,
\label{tccZ_2}
\end{eqnarray}
where $\chi =1(2,4)$ is the intersection number on $T_{2,3}$
for {\bf aa} ({\bf ab}, {\bf bb}). 

There are three major differences as compared to
section~\ref{Z-3models}: First of all, worldsheet duality gives a
constraint on the representation matrices of the orbifold group
leading to a smaller gauge group,
\mbox{$U(N_a)\stackrel{\Z_2}{\longrightarrow} U(N_a/2)\times U(N_a/2)$}
[$\stackrel{\OR}{\longrightarrow}U(N_a/2)$ for a brane which is its
own mirror brane]. Secondly, for a given wrapping number on $T_1$
there exist two different brane configurations on $T_{2,3}$,
namely those extended along the $(x^6, x^8)$- and those along the
$(x^7, x^9)$-directions, 
which cannot be identified by any element of the orientifold
group. Thus, for given $(n_a,m_a)$ we obtain the total gauge group
$U(N_a/2)^4$ [$U(N_a/2)^2$ for $m_a=0$]. 
At last, the intersection number $\chi$ on $T_{2,3}$ explicitly
enters the tadpole cancellation condition. A $\Z_2$ rotation maps each
sector onto itself whilst assigning a fixed parity $\pm 1$ to each
massless state. As one can easily read off from 
table~\ref{fermionic_groundstates}, left-handed states are
$\Z_2$-even and right-handed ones are $\Z_2$-odd in all sectors with a
non-vanishing angle on $T_1$. Therefore, not only strings stretching
between branes at non-vanishing angles on all three tori contribute to
the chiral spectrum but also those which are merely tilted on
$T_1$. This is also the reason why the intersection number $\chi$
explicitly enters the tadpole cancellation conditions~(\ref{tccZ_2}). 
Taking all this into account, the chiral part of the open string
spectrum is as listed in table~\ref{Z-2spectrum}.

The analysis of the $\Z_{4,6}$ cases goes completely analogous to
$\Z_2$. Again, there exist two independent brane configurations on
$T_{2,3}$ and a condition on the Chan-Paton matrices,
\mbox{$\text{tr}\gamma_{\Theta^{N/2}}=0$}, yielding the gauge group
$U(N_a/2)^4$ for given $(n_a,m_a)$. ``Untwisted'' strings on
$T_{2,3}$ will contribute to the chiral spectrum since $\Z_2$-even and -odd
states have opposite handedness. One additional subtlety enters the
computation of the ``$\Theta^2$-twisted'' and ``$\Theta^3$-twisted'' open
spectrum in the case of $\Z_6$ as the intersection points on $T_{2,3}$ are
permuted by the orbifold group. However, the tadpole
conditions~(\ref{tcc}) already indicate that we cannot include the
standard model gauge group \mbox{$SU(3)\times SU(2)\times U(1)$} in
$\Z_{4,6}$ without adding anti-branes. Therefore, we will not discuss
these models in detail but close this section by giving a $\Z_2$ example
which encloses \mbox{$SU(3)\times SU(2)\times U(1)$}.
\renewcommand{\arraystretch}{1.5}
\begin{table}[ht]
  \begin{center}
    \begin{equation*}
      \begin{array}{|c||c|c|} \hline
        \multicolumn{3}{|c|}{\rule[-3mm]{0mm}{8mm} \text{\bf
         Chiral fermionic spectrum for $\Z_2$}} \\ \hline\hline
        &\text{rep. of } U(\frac{N_a}{2})^4 \times U(\frac{N_b}{2})^4 & \text{mult.} 
\\ \hline\hline
aa'U & (\F_a,\F_a,1,1)_L+(1,1,\F_a,\F_a)_L & 4m_an_a\\
& \underline{(\Bar{\Anti}_a,1,1,1)}_L & 4m_a\\
& \underline{(\Bar{\Anti}_a+\Bar{\Sym}_a,1,1,1)}_L &
2m_a(n_a-1)\\\hline
aa'T & (\Bar{\F}_a,1,1,\Bar{\F}_a)_L+(1,\Bar{\F}_a,\Bar{\F}_a,1)_L & 2m_an_a\chi\\\hline
abU & (\F_a,1,1,1;1,\Bar{\F}_b,1,1)_L+(1,\F_a,1,1;\Bar{\F}_b,1,1,1)_L  
 & 2(n_am_b-n_bm_a)\\
& +(1,1,\F_a,1;1,1,1,\Bar{\F}_b)_L+(1,1,1,\F_a;1,1,\Bar{\F}_b,1)_L & \\
& (\Bar{\F}_a,1,1,1;\F_b,1,1,1)_L+(1,\Bar{\F}_a,1,1;1,\F_b,1,1)_L 
& 2(n_am_b-n_bm_a)\\
& +(1,1,\Bar{\F}_a,1;1,1,\F_b,1)_L+(1,1,1,\Bar{\F}_a;1,1,1,\F_b)_L & \\
ab'U & (\Bar{\F}_a,1,1,1;\Bar{\F}_b,1,1,1)_L+(1,\Bar{\F}_a,1,1;1,\Bar{\F}_b,1,1)_L  & 2(n_am_b+n_bm_a)\\
& +(1,1,\Bar{\F}_a,1;1,1,\Bar{\F}_b,1)_L+(1,1,1,\Bar{\F}_a;1,1,1,\Bar{\F}_b)_L & \\
& (\F_a,1,1,1;1,\F_b,1,1)_L+(1,\F_a,1,1;\F_b,1,1,1)_L & 2(n_am_b+n_bm_a)\\
& +(1,1,\F_a,1;1,1,1,\F_b)_L+(1,1,1,\F_a:1,1,\F_b,1)_L & \\\hline
abT & (\F_a,1,1,1;1,1,\Bar{\F}_b,1)_L+(1,\F_a,1,1;1,1,1,\Bar{\F}_b)_L &
(n_am_b-n_bm_a)\chi\\
& +(1,1,\F_a,1;\Bar{\F}_b,1,1,1,)_L+(1,1,1,\F_a;1,\Bar{\F}_b,1,1)_L &\\
ab'T & (\Bar{\F}_a,1,1,1;1,1,1,\Bar{\F}_b)_L+(1,\Bar{\F}_a,1,1;1,1,\Bar{\F}_b,1)_L &
(n_am_b+n_bm_a)\chi\\
& +(1,1,\Bar{\F}_a,1;1,\Bar{\F}_b,1,1,)_L+(1,1,1,\Bar{\F}_a;\Bar{\F}_b,1,1,1)_L &
\\\hline 
      \end{array}
    \end{equation*}
  \end{center}
\caption{Generic chiral spectrum for $\Z_2$. ``$U$'' labels identical
  configurations on $T_{2,3}$, ``$T$'' denotes branes which are
  perpendicular on $T_{2,3}$.}
\label{Z-2spectrum}
\end{table}

\subsubsection{Example 2, $\Z_2$}

As we choose not to include anti-branes in our analysis, the SM gauge
group can only be enclosed for the choice $\chi=1$ (cf. eq.~(\ref{tccZ_2})). Taking the
{\bf aaa} lattice and the minimal possible choice, namely
\begin{eqnarray}
N_1 = 6, \qquad (n_1,m_1)=(1,1),\nonumber\\
N_2 = 4, \qquad (n_2,m_2)=(1,0),\\
N_3 = 2, \qquad (n_3,m_3)=(4,1),\nonumber
\end{eqnarray}
we obtain the gauge group $SU(3)^4 \times SU(2)^2 \times
U(1)^{10}$. Herein, a
sublety arises from the second stack of branes. The tadpole condition
(\ref{tccZ_2}) has to be modified,
\begin{equation}\label{tccZ_2-mod}
  \frac{N_2}{2} + \sum_{a\neq2} n_aN_a =16.
\end{equation}
This is in agreement with the fact that models containing only stacks
of branes with \mbox{$(n,m)=(1,0)$} are T-dual to the ones considered
in~\cite{Gimon:1996rq} for $\Z_2$ and \cite{Blumenhagen:2000md} for
$\Z_3$ leading to $U(16)^2$ and $SO(8)$ respectively. The modified
equation~(\ref{tccZ_2-mod}) reflects the fact that the additional $\OR$
symmetry of \mbox{brane 2} removes half of the degrees of freedom. Thus,
the sector $1'2$ provides the \mbox{anti-particles} for the sector $12$ (and
similarly for $23$ and $23'$), whereas generically the sector $a'b'$
contains the \mbox{anti-particles} for the sector $ab$ (and similarly $ab'$ is
paired with $a'b$). Therefore, in the sectors
$13, 13', \ldots$ we obtain an even number of generations transforming
under the 
same gauge factors whereas in the sector $12$, there exists a single
particle in the $(\3_1,\2_2)$ of $SU(3)_1\times SU(2)_2$. However, as the
complete spectrum is symmetrically distributed among the gauge factors
which are supported by a specific brane configuration,
the total number of $(\3_i,\2_j)$ representations of all possible 
\mbox{$SU(3)_i \times SU(2)_j$} ($i=1,\ldots 4; j=1,2$) combinations
is even. This can be easily seen by examining the chiral spectrum
displayed in table~\ref{Ex_2spectrum}. The model contains (at
least) six non-anomalous $U(1)$ factors. A possible set of linear
combinations is given by
\begin{eqnarray}
 \Tilde{Q}_1 &=& Q_1 + Q_2 -3Q_7-3Q_8,\nonumber\\
 \Tilde{Q}_2 &=& Q_3 + Q_4 -3Q_9-3Q_{10},\nonumber\\
 \Tilde{Q}_3 &=& Q_1 - Q_2 -3Q_5,\label{non-an_U(1)s}\\
 \Tilde{Q}_4 &=& Q_3 - Q_4 -3Q_6,\nonumber\\
 \Tilde{Q}_5 &=& -4Q_5+Q_7-  Q_8,\nonumber\\
 \Tilde{Q}_6 &=& -4Q_6+Q_9-  Q_{10}.\nonumber
\end{eqnarray}
These charges are also listed in table~\ref{Ex_2spectrum}. The
remaining anomalous $U(1)$'s should become massive by a generalized GS
mechanism.
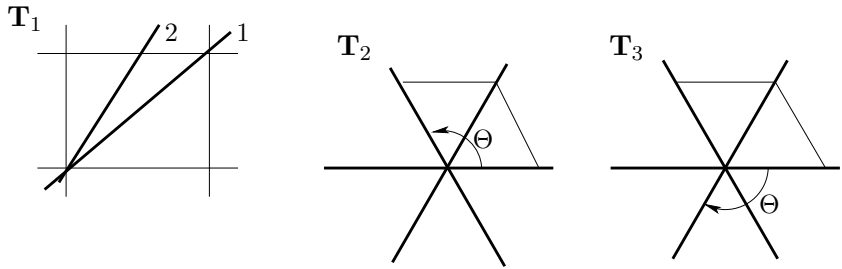
\begin{figure}[ht]
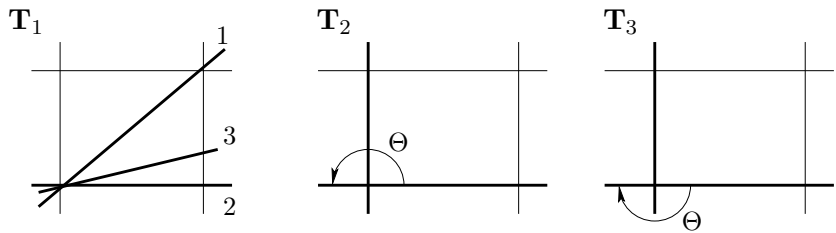
  
\begin{center}
\input examples_12.pstex_t
\end{center}
\caption{Brane configurations for the examples}
\label{examles_12}
\end{figure}

\section{Conclusions}

We have constructed type IIA theories on $\OR \times
T^2\times T^4/\Z_N$. The resulting spectra are non-supersymmetric. They
contain a bunch of chiral fermions as well as tachyons transforming in
the same representation of the gauge group. We have explicitly given
examples for $\Z_{2,3}$. Excluding anti-D-branes, for $\Z_3$ the
maximal gauge group obtainable is $SU(3)\times U(1)$ if we require the
presence of chiral fermions. The resulting spectrum is listed in
table~\ref{Ex_1spectrum}. We have also worked out an example for
$\Z_2$ which encloses the standard model gauge group as well as
several non-anomalous $U(1)$ factors. These $U(1)$'s are displayed in
equation~(\ref{non-an_U(1)s}) and the chiral spectrum is listed in
  table~\ref{Ex_2spectrum}. However, in this framework we can neither
  obtain a three generation model nor give an obvious solution to the
  hierarchy problem. Both debilities could be avoided by combining a
  more general orientifold action $\Omega\Tilde{\mathcal R}$ with
  non-vanishing background fields. Additionally, in such models
  tachyons would not necessarily carry the same gauge quantum numbers
  as the chiral fermions. We will pursue this further.
\vskip 1cm  
  
\noindent {\bf Acknowledgments} 
 
\noindent It is a pleasure to thank Ralph Blumenhagen, Boris K\"ors and
Angel Uranga for valuable discussions. Further we acknowledge
discussions with Jan O.\ Conrad, Hanno Klemm and Martin Walter.\newline
This work is supported by the European Commission RTN programs 
\mbox{HPRN-CT-2000-00131}, 00148 and 00152.


\clearpage
\begin{appendix}
\section{Computation of one-loop diagrams}
\label{app_loop}

The tadpole cancellation conditions are determined by computing those
diagrams in the loop channel which correspond to RR-exchanges in the
tree channel. As explained in section~\ref{setup}, the relevant
contributions arise from the NSNS$(-)^F$ sector for the Klein bottle,
R for the M\"obius strip and NS$(-)^F$ for the Annulus. For the
detailed calculation of the oscillator contributions in the loop
channel, we refer
to~\cite{Blumenhagen:2000ev,Forste:2001hx,Blumenhagen:2000wh}.

\subsection{Lattice contributions}
\label{app_loop_lattice}

On a torus with radii $R_{1,2}$ only momenta along the $\OR$ invariant
direction and windings perpendicular to the former ones contribute. In
the loop channel, the general contribution is given by
\begin{equation}\label{latticeContributionLoop}
  \mathcal{L}^{R_1,R_2}[\alpha,\beta](t) \equiv \left(
    \sum_{m \in \mathbb{Z}} e^{-\alpha \pi t m^2/\rho_1}\right)
  \left(\sum_{n \in \mathbb{Z}} e^{-\beta\pi t n^2 \rho_2 }\right)
\end{equation}
where $\rho_i=R^2_i/\alpha^{\prime}$. Using Poisson resummation gives
the general expression for the lattice contribution in the tree
channel
\begin{equation}\label{latticeContributionTree}
  \tilde{\mathcal{L}}^{R_1,R_2}[\alpha,\beta](t) \equiv \left(
    \sum_{m \in \mathbb{Z}} e^{-\alpha \pi t m^2\rho_1}\right)
  \left(\sum_{n \in \mathbb{Z}} e^{-\beta\pi t n^2 /\rho_2 }\right).
\end{equation}
(\ref{latticeContributionLoop}) and (\ref{latticeContributionTree})
are related via
\begin{equation}
   \mathcal{L}^{R_1,R_2}[\alpha,\beta](t) = cl
  \tilde{\mathcal{L}}^{R_1,R_2}[\frac{\kappa}{\alpha},\frac{\kappa}{\beta}]
  \left(l\right) 
\end{equation}
with $c = \frac{\kappa}{\sqrt{\alpha\beta}}\frac{R_1}{R_2}$,
$t=1/(\kappa l)$ and $\kappa=4$ $(2, 8)$ for the Klein bottle
(Annulus, M\"obius strip). The results for the different tori are
summarized in table~\ref{lattice_table}.
\renewcommand{\arraystretch}{1.5}
\begin{table}[h!]
  \begin{center}
    \begin{equation*}
      \begin{array}{|c|c|c|c|c|c|c|} \hline
        &&&\multicolumn{2}{|c|}{\Z_{2,4}} &
        \multicolumn{2}{|c|}{\Z_{3,6}} \\
        & {\text{\bf a}} & {\text{\bf b}} &
        {\text{\bf A}} & {\text{\bf B}} &
        {\text{\bf A}} & {\text{\bf B}} \\\hline\hline
        \lu^{\ku} & \lu^{R_1,R_2}[1,1] &
          \lu^{R,R}[1/\cos^2\alpha,4\sin^2\alpha] &
          \lu^{R,R}[1,1] & \lu^{R,R}[2,2] &
          \lu^{R,R}[4,3] & \lu^{R,R}[4/3,1] \\\hline
        \lt^{\ku} &\lt^{R_1,R_2}[4,4] &
          \lt^{R,R}[4\cos^2\alpha,1/\sin^2\alpha] &
          \lt^{R,R}[4,4] & \lt^{R,R}[2,2] &
          \lt^{R,R}[1,4/3] & \lt^{R,R}[3,4] \\\hline
        c^{\ku} & 4 R_1/R_2 & 2 /\tan\alpha &
          4 & 2 & 2/\sqrt{3} & 2 \sqrt{3} \\\hline\hline
        \lu^{\au} & \lu^{\au}_{\text{\bf a}} & 
          \lu^{R,R}[2/V_{\alpha}^2,2\sin^2(2\alpha)/V_{\alpha}^2] & 
          \lu^{R,R}[2,2] & \lu^{R,R}[1,1] &
          \lu^{R,R}[2,3/2] & \lu^{R,R}[2/3,1/2] \\\hline
        \lt^{\au} & \lt^{\au}_{\text{\bf a}} &
          \lt^{R,R}[V_{\alpha}^2,V_{\alpha}^2/\sin^2(2\alpha)] &
          \lt^{R,R}[1,1] & \lt^{R,R}[2,2] & 
          \lt^{R,R}[1,4/3] & \lt^{R,R}[3,4] \\\hline
        c^{\au} & V^2/(R_1R_2) & V_{\alpha}^2/\sin(2\alpha) 
          & 1 & 2 & 2/\sqrt{3} & 2 \sqrt{3} \\\hline\hline
        \lu^{\msu} & \lu^{R_1,R_2}[2,2] &
          \lu^{R,R}[1/(2\cos^2\alpha),8\sin^2\alpha]
          & \lu^{R,R}[2,2] & \lu^{R,R}[1,4] &
          \lu^{R,R}[2,6] & \lu^{R,R}[2/3,2] \\\hline
        \lt^{\msu} & \lt^{R_1,R_2}[4,4] &
          \lt^{R,R}[16\cos^2\alpha),1/\sin^2\alpha] &
          \lu^{R,R}[4,4] & \lu^{R,R}[8,2] &
          \lu^{R,R}[4,4/3] & \lu^{R,R}[12,4] \\\hline
        c^{\msu} & 4 R_1/R_2 & 4/\tan\alpha & 4 & 
          4 & 4/\sqrt{3} & 4\sqrt{3} \\\hline
        \end{array}
      \end{equation*}
    \caption{The different lattice contributions. 
          $V_{\alpha} = \sqrt{n_a^2 + m_a^2 +
          2n_am_a\cos(2\alpha)}$. The definitions of
          $\lu^{\au}_{\text{\bf a}}$ and $V$ are given in
          the text in section~\ref{amplitudes}.}
    \label{lattice_table}
  \end{center}
\end{table}
\renewcommand{\arraystretch}{1.0}

\section{Computation of tree channel diagrams}
\label{app_tree}

\subsection{Construction of crosscap states}
\label{app_tree_crosscap}

 A comprehensive introduction into the boundary state approach to D
 branes is given in~\cite{Gaberdiel:2000jr} and references
 therein. \mbox{Appendix A} 
of~\cite{Angelantonj:2000xf} contains the construction of
crosscap states in related models.

\subsubsection{Oscillator part}
\label{app_tree_crosscap_osc}
Crosscap states have to fulfill
\begin{equation}\label{app_crosscap_cond}
  \left[X^i_{L,R}(\sigma,0)-\Theta^k{\cal R}X^i_{R,L}(\sigma+\pi,0)\right] |\OR \Theta^k \rangle=0.
\end{equation}
Inserting the mode expansion 
\begin{equation}\label{app_mode-exp_X}
X^i(\sigma,
\tau)=x^i+\frac{p^i_L}{2\pi}(\tau+\sigma)+\frac{p^i_R}{2\pi}(\tau-\sigma) 
+\frac{i}{2}\sum_r\frac{1}{r}\alpha^i_r e^{-ir(\tau+\sigma)}
+\frac{i}{2}\sum_s\frac{1}{s}\Tilde{\alpha}^i_s e^{-is(\tau-\sigma)}
\end{equation}
gives the constraints in terms of bosonic oscillators
\begin{eqnarray}
(\alpha^{\mu}_n+(-1)^n\Tilde{\alpha}^{\mu}_{-n})|\OR\Theta^k,\eta\rangle=0,\nonumber\\
(\alpha^i_r +(-1)^re^{2\pi i k
  v_i}\Tilde{\alpha}^{\Bar{i}}_{-r})|\OR\Theta^k,
\eta\rangle=0,\\
(\alpha^{\Bar{i}}_r +(-1)^re^{-2\pi i k v_i}\Tilde{\alpha}^i_{-r})|\OR\Theta^k,\eta\rangle=0.\nonumber
\end{eqnarray}
The constraints for the fermionic oscillators are $(\eta = \pm 1)$
\begin{eqnarray}
(\psi^{\mu}_r+i \eta e^{-i \pi r}
\Tilde{\psi}^{\mu}_{-r})|\OR\Theta^k, \eta\rangle=0,\nonumber\\
(\psi^i_r+i \eta e^{-i \pi r}e^{2\pi i k v_i}
\Tilde{\psi}^{\Bar{i}}_{-r})|\OR\Theta^k,\eta\rangle=0,\\
(\psi^{\Bar{i}}_r+i \eta e^{-i \pi r} e^{-2\pi i k v_i}
\Tilde{\psi}^i_{-r})|\OR\Theta^k, \eta\rangle=0.\nonumber
\end{eqnarray}
A solution is provided by
\begin{eqnarray}
   |\OR\Theta^k, \eta \rangle &=& \mathcal{N}_C \exp
        \Bigl\{-\sum_n\frac{(-1)^n}{n}\alpha^{\mu}_{-n} 
        \Tilde{\alpha}^{\mu}_{-n}
         -\sum_{i \in \{1,2,3\}}\sum_n\frac{(-1)^n}{n}e^{-2\pi i k
        v_i}\alpha^{i}_{-n} \Tilde{\alpha}^{i}_{-n}\nonumber\\
      &-&\sum_{\Bar{i} \in \{\Bar{1},\Bar{2},\Bar{3}\}} 
      \sum_n\frac{(-1)^n}{n}e^{2\pi i k v_i}\alpha^{\Bar{i}}_{-n}
        \Tilde{\alpha}^{\Bar{i}}_{-n}
     -i \eta\sum_r e^{-i\pi r}\psi^{\mu}_{-r}\Tilde{\psi}^{\mu}_{-r}\\
    &-&i \eta\sum_{i \in \{1,2,3\}}\sum_r e^{-i\pi r} e^{-2\pi i k v_i} \psi^{i}_{-r}\Tilde{\psi}^{i}_{-r}
    -i \eta\sum_{\Bar{i} \in \{\Bar{1},\Bar{2},\Bar{3}\}}\sum_r e^{-i\pi r}e^{2\pi i k v_i}\psi^{\Bar{i}}_{-r}\Tilde{\psi}^{\Bar{i}}_{-r}\Bigr\}|0,\eta\rangle
\nonumber
\end{eqnarray}
where $r \in \Z (\Z+\frac{1}{2})$ in the RR (NSNS) sector.
GSO-invariant states are given by
\begin{eqnarray}
|\OR\Theta^k\rangle  &=& |\OR\Theta^k\rangle_{NSNS}+|\OR\Theta^k\rangle_{RR},\nonumber\\
|\OR\Theta^k\rangle_{NSNS}&=&|\OR\Theta^k,+\rangle_{NSNS}-|\OR\Theta^k,-\rangle_{NSNS},\label{GSOinvC}\\
|\OR\Theta^k\rangle_{RR}&=&|\OR\Theta^k,+\rangle_{RR}-i|\OR\Theta^k,-\rangle_{RR}.\nonumber
\end{eqnarray}
The total crosscap state has to be invariant w.r.t.\ the orbifold
group (i.e.\ it contains the `complete projector'):
\begin{equation}\label{crosscap}
 |C\rangle  =\sum_{k=0}^{N-1} |\OR\Theta^k\rangle.
\end{equation}

\subsubsection{Lattice part}
\label{app_tree_crosscap_lattice}
From (\ref{app_crosscap_cond}) we also obtain
\begin{equation}\label{app_cross_latt_x}
\left(x^i-e^{2\pi i k v_i}
  [x^{\Bar{i}}+\frac{1}{2}(p^{\Bar{i}}_L-p^{\Bar{i}}_R)]\right)
 |\OR\Theta^k\rangle=0
\end{equation}
by inserting the mode expansion~(\ref{app_mode-exp_X}). From this we
can read off that the crosscap state $|\OR\Theta^k\rangle$ is confined
to a line on $T_i$ which is tilted by the angle $-\pi kv_i$ w.r.t.\
the real axis.
Finally, conditions on the momenta arise:
\begin{eqnarray}
 p^{\mu}|\OR\Theta^k\rangle=0,\nonumber\\
(p^i_L+e^{2\pi i k
  v_i}p^{\Bar{i}}_R)|\OR\Theta^k\rangle=0,\label{app_cross_latt_p}\\ 
(p^i_R+e^{2\pi i k v_i}p^{\Bar{i}}_L)|\OR\Theta^k\rangle=0.\nonumber    
\end{eqnarray}
Inserting $p_{L,R}=P\pm\alpha'W$ for the compact momenta,
(\ref{app_cross_latt_p}) indicates that on each $T_i$, there are
Kaluza-Klein momenta perpendicular and windings parallel to the
position of the crosscap state.  

\subsection{Boundary states}
\label{app_tree_boundary}

Similarly to a crosscap state, the boundary state for a D-brane at angle
$\pi \varphi$ on 
$T_1$ w.r.t.\ the $x^4$ axis is given by
\begin{eqnarray}
   |\varphi,\Theta^k; \eta \rangle = &\mathcal{N}_{B}&\exp
        \Bigl\{-\sum_n\frac{1}{n}\alpha^{\mu}_{-n} \Tilde{\alpha}^{\mu}_{-n}
         -\sum_n\frac{1}{n}e^{2\pi i \varphi}\alpha^{1}_{-n}
         \Tilde{\alpha}^{1}_{-n}
        -\sum_{i \in \{2,3\}}\sum_n\frac{1}{n}e^{-2\pi i k v_i}\alpha^{i}_{-n}
         \Tilde{\alpha}^{i}_{-n}\nonumber\\
         &-&\sum_n\frac{1}{n}e^{-2\pi i \varphi}\alpha^{\Bar{1}}_{-n} 
         \Tilde{\alpha}^{\Bar{1}}_{-n}
         -\sum_{\Bar{i} \in \{\Bar{2},\Bar{3}\}}\sum_n\frac{1}{n}e^{2\pi i k
           v_i}\alpha^{\Bar{i}}_{-n} 
         \Tilde{\alpha}^{\Bar{i}}_{-n}\nonumber\\
         &-&i \eta\sum_r\psi^{\mu}_{-r}\Tilde{\psi}^{\mu}_{-r}
         -i \eta\sum_r  e^{2\pi i \varphi} \psi^{1}_{-r}\Tilde{\psi}^{1}_{-r}
         -i \eta \sum_{i \in \{2,3\}} \sum_r  e^{-2\pi i k v_i} \psi^{i}_{-r}
         \Tilde{\psi}^{i}_{-r}\\
       &-&i \eta\sum_re^{-2\pi i \varphi
         }\psi^{\Bar{1}}_{-r}\Tilde{\psi}^{\Bar{1}}_{-r}
         -i \eta\sum_{\Bar{i} \in \{\Bar{2},\Bar{3}\}}\sum_re^{2\pi i k
           v_i}\psi^{\Bar{i}}_{-r}\Tilde{\psi}^{\Bar{i}}_{-r}\Bigr\}|0,
         \eta\rangle \nonumber  
\end{eqnarray}
Using the analogous equations to (\ref{GSOinvC}), the GSO-ivariant
boundary state is given by
\begin{equation}\label{boundary6}
|B_{\varphi}\rangle =\sum_{k=0}^{N-1} |\varphi,\Theta^k\rangle.
\end{equation}
Discrete momenta exist in the compact directions perpendicular to the
position of the boundary state while windings are parallel.

\subsection{Tree channel amplitudes}
\label{app_tree_amplitudes}

Using equation (\ref{crosscap}) and (\ref{boundary6}), the tree
channel amplitudes read
\begin{eqnarray}
\Tilde{\KB}_{total}&=& \Tilde{\KB}_{RR}+ \Tilde{\KB}_{NSNS}
=\int_0^{\infty}d l\langle C|e^{-2\pi lH_{cl}}|C\rangle\nonumber\\
\Tilde{\AN}_{total}&=&\Tilde{\AN}_{RR}+\Tilde{\AN}_{NSNS} 
=\int_0^{\infty}d l \sum_{\varphi,\varphi'} \langle B_{\varphi}|e^{-2\pi
  lH_{cl}}|B_{\varphi'}\rangle\\
\Tilde{\MS}_{total}&=&\Tilde{\MS}_{RR}+\Tilde{\MS}_{NSNS}
=\int_0^{\infty}d l \sum_{\varphi}\left(\langle C|e^{-2\pi lH_{cl}}
  |B_{\varphi}\rangle + h.c.\right)\nonumber
\end{eqnarray}
As we only need to compute the RR exchange, we will use the abbreviation
$\Tilde{\KB}\equiv \Tilde{\KB}_{RR}$ etc. 
The normalizations $\mathcal{N}_{C}, \mathcal{N}_{B}$ are determined
from the Klein bottle and Annulus amplitude via worldsheet duality.
The following equation holds
\begin{equation}
\renewcommand{\arraystretch}{1.3}
\langle\OR\Theta^k|e^{-2\pi lH_{cl}}|\OR\Theta^{k'}\rangle=
\left\{\begin{array}{lc}
\mathcal{N}_{C}^2\kt^{(0)}\lt_1\lt_2\lt_3 &\mbox{for }k=k'\\
\prod_{i=1}^2\left[-2\sin(\pi v_i(k-k'))\right]
\mathcal{N}_{C}^2\kt^{(k-k')}\lt_1 & \mbox{for }k\neq k'
\end{array}\right.
\end{equation}
where $\kt^{(k-k')}\equiv\kt^{(k-k')}(l)$ contains the oscillator
contribution (for  
notation see~\cite{Blumenhagen:2000ev,Forste:2001hx}) and $\lt_i\equiv\lt_i(l)$
denotes the lattice contribution for the torus $T_i$. 
For $\Z_{2,3}$, all $|\OR\Theta^k\rangle$ are extended along (diagonal
to) the axes of $T_{2,3}$ in the case of the {\bf A} ({\bf B}) lattice.
Since branes lie on top of O-planes on $T_{2,3}$ in our
models, the positions of the O-planes can be read off from
figure~\ref{examles_12}. All $|\OR\Theta^k\rangle$ have the same
relative orientation w.r.t.\ the tori $T_{2,3}$,
hence they all provide the same lattice
contribution. All choices {\bf AA}, {\bf AB}, {\bf BB} lead to
consistent models. 
The situation is different for $\Z_{4,6}$. In these models, all
$|\OR\Theta^{2k}\rangle$ have the same orientation w.r.t.\ the lattice
while all $|\OR\Theta^{2k+1}\rangle$ have the other possible
one. $|\OR\Theta^{2k}\rangle$ on the lattice {\bf A} gives the same
contribution as $|\OR\Theta^{2k+1}\rangle$ on {\bf B} and vice versa,
\begin{equation}
\renewcommand{\arraystretch}{1.3}
 \langle\OR\Theta^{2k}|e^{-2\pi lH_{cl}}|\OR\Theta^{2k}\rangle
+\langle\OR\Theta^{2k+1}|e^{-2\pi lH_{cl}}|\OR\Theta^{2k+1}\rangle
\longrightarrow
\left\{\begin{array}{cc}
 2\lt_{\bf A}\lt_{\bf B} & \text{for \bf AB}\\
\lt_{\bf A}^2+\lt_{\bf B}^2 & \text{for {\bf AA}/{\bf BB}}
\end{array}\right.
\label{worldsheet_cons}
\end{equation}
By modular transformation from the loop channel, one recovers the
correct result for {\bf AB}. But for {\bf AA} or {\bf BB}, the loop
channel amplitude gives  
$4\lt_{\bf A}^2+\lt_{\bf B}^2$ for $\Z_4$ ($\lt_{\bf A}^2+9\lt_{\bf
  B}^2$ for $\Z_6$) as can be read off from
table~\ref{lattice_table}. This means that only the {\bf AB}-lattice
is consistent with worldsheet duality.

\section{Spectra}
\label{app_spectra}

\subsection{Closed string spectra}
\label{app_spectra_closed}

\begin{equation*}
  \renewcommand{\arraystretch}{1.3}
  \begin{array}{cc}
    \begin{array}{|c||c|c|c|} \hline
          \multicolumn{4}{|c|}{\rule[-3mm]{0mm}{8mm} \text{\bf
              Closed string spectrum for $\Z_2$}} \\ \hline\hline
          \text{twist-sector} & \text{\bf AA} & \text{\bf AB} &
          \text{\bf BB}\\ \hline\hline
          \mbox{untwisted} &  \multicolumn{3}{|c|}{
            \mbox{SUGRA} + 11\mbox{C}+4\mbox{V}}\\ \hline
          \theta_1 & 32\mbox{C} & 28\mbox{C} + 4\mbox{V} & 26\mbox{C} +
          6\mbox{V}\\ \hline\hline  
          \multicolumn{4}{|c|}{\rule[-3mm]{0mm}{8mm} \text{\bf
              Closed spectrum of $\Z_3$}} \\ \hline\hline
          \mbox{untwisted} & \multicolumn{3}{|c|}{\mbox{SUGRA} + 8
            \mbox{C}+ 3\mbox{V}} \\ \hline 
          \theta_1 + \theta_1^2 & 28\mbox{C} + 8\mbox{V} & 30\mbox{C} +
          6\mbox{V} & 36\mbox{C} \\ \hline 
        \end{array}
        &
        \begin{array}{|c||c|} \hline
          \multicolumn{2}{|c|}{\rule[-3mm]{0mm}{8mm} \text{\bf
              Closed string spectrum for $\Z_4$}} \\ \hline\hline
          \text{lattice} & \text{\bf AB}  \\ \hline\hline
          \mbox{untwisted} & \mbox{SUGRA} + 8\mbox{C}+3\mbox{V}\\ \hline
          \theta_1 + \theta_1^3 & 16C\\ \hline
          \theta_1^2  & 19C + 1V   \\ \hline\hline
          \multicolumn{2}{|c|}{\rule[-3mm]{0mm}{8mm} \text{\bf
              Closed string spectrum for $\Z_6$}} \\ \hline\hline
          \mbox{untwisted}&\mbox{SUGRA}+8\mbox{C}+3\mbox{V}\\ \hline
          \theta_1 + \theta_1^5 &  4\mbox{C} \\ \hline
          \theta_1^2 + \theta_1^4 & 18\mbox{C} +2\mbox{V} \\ \hline
          \theta_1^3  & 11\mbox{C} + 1\mbox{V} \\ \hline
    \end{array}
  \end{array}
\end{equation*}

\subsection{Open string spectra}
\label{app_spectra_open}

The multiplicity of open string states is determined by the number of
intersections of brane $a$ and $b$ on $T_1$ in terms of their
wrapping numbers,
\begin{equation}
  I_{ab}=n_{a}m_{b}-m_{a}n_{b},
\end{equation}
as well as by the number of intersections $\chi$ on $T_{2,3}$. An {\bf
  A} type torus contributes a factor of one, whereas a {\bf B} type
  torus contributes a factor of 2 (3) in the case of $\Z_{2,4}$ ($\Z_{3,6}$).

For mirror branes, we have to distinguish between $\OR$ invariant
intersections and those which form pairs under $\OR$:
\begin{align}
  I_{a'a}^{\OR}&\equiv\left\{\begin{array}{cc}
      2m_{a}& {\bf a}\\
      m_{a}-n_{a} & {\bf b} 
    \end{array}\right.\\
  I_{a'a}-I_{a'a}^{\OR}&=\left\{
    \begin{array}{cc}
      2 m_{a}(n_{a}-1) & {\bf a}\\
      \left[ m_{a}(m_{a}-1)-n_{a}(n_{a}-1)\right] & {\bf b} 
    \end{array}\right.
\end{align}

\subsubsection{Fermionic states}
\label{app_spectra_open_fermion}

The sectors where the branes are parallel on the first torus
are non-chiral. In the sector with $\Delta\varphi \neq
0$ and $\frac{k}{N}\neq 0$, the left-handed fermion is massless while the
would-be right-handed one becomes massive with
\mbox{$\alpha^{\prime}m^2 = \Delta\varphi$}. For $\Z_{2,4,6}$,
additional chiral fermions arise from strings stretching between
mirror branes with $\frac{k}{N}=0$ because the left handed R-states are
even under the $\Z_2$ symmetry while the right handed ones are odd and
thus are subject to a different orbifold projection.
\renewcommand{\arraystretch}{1.1}
\begin{table}[ht]
  \begin{center}
    \begin{equation*}
      \begin{array}{|c|c||l|c|c||c|} \hline
        \multicolumn{6}{|c|}{\rule[-3mm]{0mm}{8mm} \text{\bf
            Fermionic states on  $T^2 \times T^4/\Z_N$}} \\ \hline\hline
        \text{on }T^2 & \text{on }T^4/\Z_2 & \text{state} &\text{mass}&
        \text{chirality} & \Z_2
\\ \hline\hline
\Delta\varphi=0& \frac{k}{N}=0 & |0\rangle_R & 0 & L & + \\\hline
&& \psi^0_0 \psi^1_0 |0\rangle_R & 0 & R & + \\\hline 
&& \psi^0_0 \psi^2_0 |0\rangle_R & 0 & R & - \\\hline 
&& \psi^0_0 \psi^3_0 |0\rangle_R & 0 & R & - \\\hline 
&& \psi^1_0 \psi^2_0 |0\rangle_R & 0 & L & - \\\hline 
&& \psi^1_0 \psi^3_0 |0\rangle_R & 0 & L & - \\\hline
&& \psi^2_0 \psi^3_0 |0\rangle_R & 0 & L & + \\\hline 
&& \psi^0_0 \psi^1_0 \psi^2_0 \psi^3_0 |0\rangle_R & 0 & R & + \\\hline\hline  
\Delta\varphi=0& 0<\frac{k}{N}\leqslant \frac{1}{2} 
& |0\rangle_R & 0 & L & + \\\hline
&& \psi^0_0 \psi^1_0 |0\rangle_R & 0 & R & + \\\hline\hline
\Delta\varphi \neq 0 & \frac{k}{N}=0 & |0\rangle_R & 0 & L & + \\\hline
&& \psi^0_0 \psi^2_0 |0\rangle_R & 0 & R & - \\\hline 
&& \psi^0_0 \psi^3_0 |0\rangle_R & 0 & R & - \\\hline 
&& \psi^2_0 \psi^3_0 |0\rangle_R & 0 & L & + \\\hline\hline 
\Delta\varphi \neq 0 &  0<\frac{k}{N}\leqslant \frac{1}{2} & |0\rangle_R & 0
& L & + \\\hline
&& \psi^0_0 \psi^1_{-\Delta\varphi} |0\rangle_R & \Delta\varphi & (R) & (+) \\\hline 
      \end{array}
    \end{equation*}
\caption{Fermionic ground states for the open string sector}
\label{fermionic_groundstates}
  \end{center}
\end{table}
\clearpage
\renewcommand{\arraystretch}{1.3}
\begin{table}[h!]
  \begin{center}
    \begin{equation*}
      \begin{array}{|c||c|c||c|c|c|c||c|c||c|c|c|c||c|c|c|c|c|c|} \hline
        \multicolumn{19}{|c|}{\rule[-3mm]{0mm}{8mm} \text{\bf
         Chiral spectrum, Ex. 2, Part 1 }} \\ \hline\hline
        &\text{rep.} & \!\!\!\text{mult}\!\!\! & Q_1 & Q_2 & Q_3 & Q_4 & Q_5 & Q_6 & Q_7 & Q_8 & Q_9 & Q_{10}  & \Tilde{Q}_1 & \Tilde{Q}_2
         & \Tilde{Q}_3 & \Tilde{Q}_4 & \Tilde{Q}_5 & \Tilde{Q}_6
\\ \hline\hline
\!\!11'U\!\! & \!\!(\3,\3,1,1;1,1)\!\! & 4 & 1 & 1 & 0 & 0 & 0 & 0 & 0 & 0 & 0 & 0 
& 2 & 0 & 0 & 0 & 0 & 0\\
     & \!\!(1,1,\3,\3;1,1)\!\! & 4 & 0 & 0 & 1 & 1 & 0 & 0 & 0 & 0 & 0 & 0 
& 0 & 2 & 0 & 0 & 0 & 0\\
      & \!\!(\3,1,1,1;1,1)\!\! & 4 &-2 & 0 & 0 & 0 & 0 & 0 & 0 & 0 & 0 & 0 
&-2 & 0 &-2 & 0 & 0 & 0\\
      & \!\!(1,\3,1,1;1,1)\!\! & 4 & 0 &-2 & 0 & 0 & 0 & 0 & 0 & 0 & 0 & 0 
&-2 & 0 & 2 & 0 & 0 & 0\\
      & \!\!(1,1,\3,1;1,1)\!\! & 4 & 0 & 0 &-2 & 0 & 0 & 0 & 0 & 0 & 0 & 0 
& 0 &-2 & 0 &-2 & 0 & 0\\
      & \!\!(1,1,1,\3;1,1)\!\! & 4 & 0 & 0 & 0 &-2 & 0 & 0 & 0 & 0 & 0 & 0 
& 0 &-2 & 0 & 2 & 0 & 0\\\hline
\!\!11'T\!\! & \!\!(\Bar{\3},1,1,\Bar{\3};1,1)\!\! & 2 &-1 & 0 & 0 &-1 & 0 & 0 & 0 & 0
& 0 & 0 &-1 &-1 &-1 & 1 & 0 &0\\
     & \!\!(1,\Bar{\3},\Bar{\3},1;1,1)\!\! & 2 & 0 &-1 &-1 & 0 & 0 & 0 & 0 & 0 
& 0 & 0 &-1 &-1 & 1 &-1 & 0 &0  
\\\hline
\!\!12U\!\!  & \!\!(\3,1,1,1;\Bar{\2},1)\!\! & 2 & 1 & 0 & 0 & 0 &-1 & 0 & 0 & 0 & 0 &0
& 1 & 0 & 4 & 0 & 4 & 0\\
&\!\!(\Bar{\3},1,1,1;\Bar{\2},1)\!\!& 2 &-1 & 0 & 0 & 0 &-1 & 0 & 0 & 0 & 0 &0
&-1 & 0 & 2 & 0 & 4 & 0\\
            &\!\!(1,\3,1,1;\2,1)\!\!& 2 & 0 & 1 & 0 & 0 & 1 & 0 & 0 & 0 & 0 &0
& 1 & 0 &-4 & 0 &-4 & 0\\
      &\!\!(1,\Bar{\3},1,1;\2,1)\!\!& 2 & 0 &-1 & 0 & 0 & 1 & 0 & 0 & 0 & 0 &0
&-1 & 0 &-2 & 0 & -4 & 0\\  
     & \!\!(1,1,\3,1;1,\Bar{\2})\!\!& 2 & 0 & 0 & 1 & 0 & 0 &-1 & 0 & 0 & 0 &
     0
& 0 & 1 & 0 & 4 & 0 & 4\\ 
&\!\!(1,1,\Bar{\3},1;1,\Bar{\2})\!\!& 2 & 0 & 0 &-1 & 0 & 0 &-1 & 0 & 0 & 0 &
0
& 0 &-1 & 0 & 2 & 0 & 4\\
           &\!\!(1,1,1,\3;1,\2)\!\! & 2 & 0 & 0 & 0 & 1 & 0 & 1 & 0 & 0 & 0 &
           0
& 0 & 1 & 0 &-4 & 0 &-4\\ 
      &\!\!(1,1,1,\Bar{\3};1,\2)\!\!& 2 & 0 & 0 & 0 &-1 & 0 & 1 & 0 & 0 & 0 &
      0
& 0 &-1 & 0 &-2 & 0 & -4\\\hline
\!\!12T\!\!  &\!\!(\Bar{\3},1,1,1;1,\2)\!\! & 1 &-1 & 0 & 0 & 0 & 0 & 1 & 0 & 0 & 0 &
0 &-1 & 0 &-1 &-3 & 0 &-4\\
&\!\!(1,\Bar{\3},1,1;1,\Bar{\2})\!\!& 1 & 0 &-1 & 0 & 0 & 0 &-1 & 0 & 0 & 0 &
0 &-1 & 0 & 1 & 3 & 0 & 4\\
      &\!\!(1,1,\Bar{\3},1;\2,1)\!\!& 1 & 0 & 0 &-1 & 0 & 1 & 0 & 0 & 0 & 0 &
      0
& 0 &-1 &-3 &-1 &-4 & 0\\
&\!\!(1,1,1,\Bar{\3};\Bar{\2},1)\!\!& 1 & 0 & 0 & 0 &-1 &-1 & 0 & 0 & 0 & 0 &
0
& 0 &-1 & 3 & 1 & 4 & 0\\\hline
\!\!13U\!\!  & \!\!(\3,1,1,1;1,1)\!\! & 6 & 1 & 0 & 0 & 0 & 0 & 0 &-1 & 0 & 0 & 0
& 4 & 0 & 1 & 0 &-1 & 0\\
 &\!\!(\Bar{\3},1,1,1;1,1)\!\!& 6 &-1 & 0 & 0 & 0 & 0 & 0 & 0 & 1 & 0 & 0
&-4 & 0 &-1 & 0 &-1 & 0\\
   & \!\!(1,\3,1,1;1,1)\!\!  & 6 & 0 & 1 & 0 & 0 & 0 & 0 & 0 &-1 & 0 & 0
& 4 & 0 &-1 & 0 & 1 & 0\\
&\!\!(1,\Bar{\3},1,1;1,1)\!\!& 6 & 0 &-1 & 0 & 0 & 0 & 0 & 1 & 0 & 0 & 0
&-4 & 0 & 1 & 0 & 1 & 0\\
   & \!\!(1,1,\3,1;1,1)\!\!  & 6 & 0 & 0 & 1 & 0 & 0 & 0 & 0 & 0 &-1 & 0
& 0 & 4 & 0 & 1 & 0 &-1\\
&\!\!(1,1,\Bar{\3},1;1,1)\!\!& 6 & 0 & 0 &-1 & 0 & 0 & 0 & 0 & 0 & 0 & 1
& 0 &-4 & 0 &-1 & 0 &-1\\
    & \!\!(1,1,1,\3;1,1)\!\! & 6 & 0 & 0 & 0 & 1 & 0 & 0 & 0 & 0 & 0 &-1
& 0 & 4 & 0 &-1 & 0 & 1\\
&\!\!(1,1,1,\Bar{\3};1,1)\!\!& 6 & 0 & 0 & 0 &-1 & 0 & 0 & 0 & 0 & 1 & 0
& 0 &-4 & 0 & 1 & 0 & 1\\
\!\!13'U\!\!  &\!\!(\3,1,1,1;1,1)\!\!& 10 & 1 & 0 & 0 & 0 & 0 & 0 & 0 & 1 & 0 & 0
&-2 & 0 & 1 & 0 &-1 & 0\\
&\!\!(\Bar{\3},1,1,1;1,1)\!\!& 10 &-1 & 0 & 0 & 0 & 0 & 0 &-1 & 0 & 0 & 0
& 2 & 0 &-1 & 0 &-1 & 0\\
  & \!\!(1,\3,1,1;1,1)\!\!   & 10 & 0 & 1 & 0 & 0 & 0 & 0 & 1 & 0 & 0 & 0
&-2 & 0 &-1 & 0 & 1 & 0\\
&\!\!(1,\Bar{\3},1,1;1,1)\!\!& 10 & 0 &-1 & 0 & 0 & 0 & 0 & 0 &-1 & 0 & 0
& 2 & 0 & 1 & 0 & 1 & 0\\
   & \!\!(1,1,\3,1;1,1)\!\!  & 10 & 0 & 0 & 1 & 0 & 0 & 0 & 0 & 0 & 0 & 1
& 0 &-2 & 0 & 1 & 0 &-1\\
&\!\!(1,1,\Bar{\3},1;1,1)\!\!& 10 & 0 & 0 &-1 & 0 & 0 & 0 & 0 & 0 &-1 & 0
& 0 & 2 & 0 &-1 & 0 &-1\\
    & \!\!(1,1,1,\3;1,1)\!\! & 10 & 0 & 0 & 0 & 1 & 0 & 0 & 0 & 0 & 1 & 0
& 0 &-2 & 0 &-1 & 0 & 1\\
&\!\!(1,1,1,\Bar{\3};1,1)\!\!& 10 & 0 & 0 & 0 &-1 & 0 & 0 & 0 & 0 & 0 &-1
& 0 & 2 & 0 & 1 & 0 & 1\\\hline
      \end{array}
    \end{equation*}
  \end{center}
\end{table}
\clearpage
\renewcommand{\arraystretch}{1.3}
\begin{table}[h!]
  \begin{center}
    \begin{equation*}
      \begin{array}{|c||c|c||c|c|c|c||c|c||c|c|c|c||c|c|c|c|c|c|} \hline
        \multicolumn{19}{|c|}{\rule[-3mm]{0mm}{8mm} \text{\bf
         Chiral spectrum, Ex. 2, Part 2 }} \\ \hline\hline
        &\text{rep.} & \!\!\!\text{mult}\!\!\! & Q_1 & Q_2 & Q_3 & Q_4 & Q_5 & Q_6 & Q_7 & Q_8 & Q_9 & Q_{10}  & \Tilde{Q}_1 & \Tilde{Q}_2
         & \Tilde{Q}_3 & \Tilde{Q}_4 & \Tilde{Q}_5 & \Tilde{Q}_6
\\ \hline\hline
\!\!13T\!\! &\!\!(\Bar{\3},1,1,1;1,1)\!\!& 3 &-1 & 0 & 0 & 0 & 0 & 0 & 0 & 0 & 1 & 0
&-1 &-3 &-1 & 0 & 0 & 1\\
    &\!\!(1,\Bar{\3},1,1;1,1)\!\!& 3 & 0 &-1 & 0 & 0 & 0 & 0 & 0 & 0 & 0 & 1
&-1 &-3 & 1 & 0 & 0 &-1\\
    &\!\!(1,1,\Bar{\3},1;1,1)\!\!& 3 & 0 & 0 &-1 & 0 & 0 & 0 & 1 & 0 & 0 & 0
&-3 &-1 & 0 &-1 & 1 & 0\\ 
    &\!\!(1,1,1,\Bar{\3};1,1)\!\!& 3 & 0 & 0 & 0 &-1 & 0 & 0 & 0 & 1 & 0 & 0
&-3 &-1 & 0 & 1 &-1 & 0\\ 
\!\!13'T\!\! &\!\!(\Bar{\3},1,1,1;1,1)\!\!& 5 &-1 & 0 & 0 & 0 & 0 & 0 & 0 & 0 & 0 &-1
&-1 & 3 &-1 & 0 & 0 & 1\\
     &\!\!(1,\Bar{\3},1,1;1,1)\!\!& 5 & 0 &-1 & 0 & 0 & 0 & 0 & 0 & 0 &-1 & 0
&-1 & 3 & 1 & 0 & 0 &-1\\
     &\!\!(1,1,\Bar{\3},1;1,1)\!\!& 5 & 0 & 0 &-1 & 0 & 0 & 0 & 0 &-1 & 0 & 0
& 3 &-1 & 0 &-1 & 1 & 0\\
     &\!\!(1,1,1,\Bar{\3};1,1)\!\!& 5 & 0 & 0 & 0 &-1 & 0 & 0 &-1 & 0 & 0 & 0
& 3 &-1 & 0 & 1 &-1 & 0\\\hline 
\!\!23U\!\!   &\!\!(1,1,1,1;\2,1)\!\!& 2 & 0 & 0 & 0 & 0 & 1 & 0 & 1 & 0 & 0 & 0
&-3 & 0 &-3 & 0 &-3 & 0\\
      &\!\!(1,1,1,1;\2,1)\!\!& 2 & 0 & 0 & 0 & 0 & 1 & 0 &-1 & 0 & 0 & 0
& 3 & 0 &-3 & 0 & -5 & 0\\
&\!\!(1,1,1,1;\Bar{\2},1)\!\!& 2 & 0 & 0 & 0 & 0 &-1 & 0 & 0 & 1 & 0 & 0
&-3 & 0 & 3 & 0 & 3 & 0\\
&\!\!(1,1,1,1;\Bar{\2},1)\!\!& 2 & 0 & 0 & 0 & 0 &-1 & 0 & 0 &-1 & 0 & 0
& 3 & 0 & 3 & 0 & 5 & 0\\
      &\!\!(1,1,1,1;1,\2)\!\!& 2 & 0 & 0 & 0 & 0 & 0 & 1 & 0 & 0 & 1 & 0
& 0 &-3 & 0 &-3 & 0 &-3\\
      &\!\!(1,1,1,1;1,\2)\!\!& 2 & 0 & 0 & 0 & 0 & 0 & 1 & 0 & 0 &-1 & 0
& 0 & 3 & 0 &-3 & 0 & -5\\
&\!\!(1,1,1,1;1,\Bar{\2})\!\!& 2 & 0 & 0 & 0 & 0 & 0 &-1 & 0 & 0 & 0 & 1
& 0 &-3 & 0 & 3 & 0 & 3\\ 
&\!\!(1,1,1,1;1,\Bar{\2})\!\!& 2 & 0 & 0 & 0 & 0 & 0 &-1 & 0 & 0 & 0 &-1
& 0 & 3 & 0 & 3 & 0 & 5\\\hline
\!\!23T\!\!  & \!\!(1,1,1,1;\2,1)\!\!& 1 & 0 & 0 & 0 & 0 & 1 & 0 & 0 & 0 &-1 & 0
& 0 & 3 &-3 & 0 &-4 &-1\\ 
&\!\!(1,1,1,1;\Bar{\2},1)\!\!& 1 & 0 & 0 & 0 & 0 &-1 & 0 & 0 & 0 & 0 &-1
& 0 & 3 & 3 & 0 & 4 & 1\\ 
      &\!\!(1,1,1,1;1,\2)\!\!& 1 & 0 & 0 & 0 & 0 & 0 & 1 &-1 & 0 & 0 & 0
& 3 & 0 & 0 &-3 &-1 &-4\\ 
&\!\!(1,1,1,1;1,\Bar{\2})\!\!& 1 & 0 & 0 & 0 & 0 & 0 &-1 & 0 &-1 & 0 & 0
& 3 & 0 & 0 & 3 & 1 & 4\\\hline
\!\!33'U\!\! &\!\!(1,1,1,1;1,1)\!\!& 16 & 0 & 0 & 0 & 0 & 0 & 0 & 1 & 1 & 0 & 0
&-6 & 0 & 0 & 0 & 0 & 0\\
     &\!\!(1,1,1,1;1,1)\!\!& 16 & 0 & 0 & 0 & 0 & 0 & 0 & 0 & 0 & 1 & 1
& 0 &-6 & 0 & 0 & 0 & 0\\
     &\!\!(1,1,1,1;1,1)\!\!& 6 & 0 & 0 & 0 & 0 & 0 & 0 &-2 & 0 & 0 & 0
& 6 & 0 & 0 & 0 &-2 & 0\\
     &\!\!(1,1,1,1;1,1)\!\!& 6 & 0 & 0 & 0 & 0 & 0 & 0 & 0 &-2 & 0 & 0
& 6 & 0 & 0 & 0 & 2 & 0\\
     &\!\!(1,1,1,1;1,1)\!\!& 6 & 0 & 0 & 0 & 0 & 0 & 0 & 0 & 0 &-2 & 0
& 0 & 6 & 0 & 0 & 0 &-2\\
     &\!\!(1,1,1,1;1,1)\!\!& 6 & 0 & 0 & 0 & 0 & 0 & 0 & 0 & 0 & 0 &-2
& 0 & 6 & 0 & 0 & 0 & 2\\\hline
\!\!33'T\!\! &\!\!(1,1,1,1;1,1)\!\!& 8 & 0 & 0 & 0 & 0 & 0 & 0 &-1 & 0 & 0 &-1
& 3 & 3 & 0 & 0 &-1 & 1\\
     &\!\!(1,1,1,1;1,1)\!\!& 8 & 0 & 0 & 0 & 0 & 0 & 0 & 0 &-1 &-1 & 0
& 3 & 3 & 0 & 0 & 1 &-1\\\hline
      \end{array}
    \end{equation*}
  \end{center}
 \caption{Chiral fermionic spectrum for $\Z_2$ with $(n_1,m_1)=(1,1)$,
   $(n_2,m_2)=(1,0)$, $(n_3,m_3)=(4,1)$ and lattice {\bf aaa}. The
   resulting gauge group is $SU(3)^4 \times SU(2)^2 \times
   U(1)^{10}$.} 
\label{Ex_2spectrum}
\end{table}  
\end{appendix}

\end{document}

%% file: o_planes_4.pstex_t
\begin{picture}(0,0)%
\includegraphics{o_planes_4.pstex}%
\end{picture}%
\setlength{\unitlength}{2368sp}%
\begingroup\makeatletter\ifx\SetFigFont\undefined%
\gdef\SetFigFont#1#2#3#4#5{%
  \reset@font\fontsize{#1}{#2pt}%
  \fontfamily{#3}\fontseries{#4}\fontshape{#5}%
  \selectfont}%
\fi\endgroup%
\begin{picture}(9549,3453)(364,-6298)
\put(5251,-3586){\makebox(0,0)[lb]{\smash{\SetFigFont{12}{14.4}{\rmdefault}{\mddefault}{\updefault}\special{ps: gsave 0 0 0 setrgbcolor}$\Theta$\special{ps: grestore}}}}
\put(9001,-4861){\makebox(0,0)[lb]{\smash{\SetFigFont{12}{14.4}{\rmdefault}{\mddefault}{\updefault}\special{ps: gsave 0 0 0 setrgbcolor}$\Theta$\special{ps: grestore}}}}
\put(6901,-3061){\makebox(0,0)[lb]{\smash{\SetFigFont{12}{14.4}{\rmdefault}{\mddefault}{\updefault}\special{ps: gsave 0 0 0 setrgbcolor}$\mbox{\bf T}_{\mbox{\scriptsize \bf 3}}$\special{ps: grestore}}}}
\put(3676,-3061){\makebox(0,0)[lb]{\smash{\SetFigFont{12}{14.4}{\rmdefault}{\mddefault}{\updefault}\special{ps: gsave 0 0 0 setrgbcolor}$\mbox{\bf T}_{\mbox{\scriptsize \bf 2}}$\special{ps: grestore}}}}
\put(451,-3061){\makebox(0,0)[lb]{\smash{\SetFigFont{12}{14.4}{\rmdefault}{\mddefault}{\updefault}\special{ps: gsave 0 0 0 setrgbcolor}$\mbox{\bf T}_{\mbox{\scriptsize \bf 1}}$\special{ps: grestore}}}}
\end{picture}

%% file: torus_1.pstex_t
\begin{picture}(0,0)%
\includegraphics{torus_1.pstex}%
\end{picture}%
\setlength{\unitlength}{1776sp}%
\begingroup\makeatletter\ifx\SetFigFont\undefined%
\gdef\SetFigFont#1#2#3#4#5{%
  \reset@font\fontsize{#1}{#2pt}%
  \fontfamily{#3}\fontseries{#4}\fontshape{#5}%
  \selectfont}%
\fi\endgroup%
\begin{picture}(12174,4674)(439,-5098)
\put(2251,-1336){\makebox(0,0)[lb]{\smash{\SetFigFont{14}{16.8}{\rmdefault}{\mddefault}{\updefault}\special{ps: gsave 0 0 0 setrgbcolor}{\bf a}\special{ps: grestore}}}}
\put(7651,-2161){\makebox(0,0)[lb]{\smash{\SetFigFont{14}{16.8}{\rmdefault}{\mddefault}{\updefault}\special{ps: gsave 0 0 0 setrgbcolor}$R$\special{ps: grestore}}}}
\put(8176,-1261){\makebox(0,0)[lb]{\smash{\SetFigFont{14}{16.8}{\rmdefault}{\mddefault}{\updefault}\special{ps: gsave 0 0 0 setrgbcolor}{\bf b}\special{ps: grestore}}}}
\put(7651,-3811){\makebox(0,0)[lb]{\smash{\SetFigFont{14}{16.8}{\rmdefault}{\mddefault}{\updefault}\special{ps: gsave 0 0 0 setrgbcolor}$R$\special{ps: grestore}}}}
\put(601,-3511){\makebox(0,0)[lb]{\smash{\SetFigFont{14}{16.8}{\rmdefault}{\mddefault}{\updefault}\special{ps: gsave 0 0 0 setrgbcolor}$R_2$\special{ps: grestore}}}}
\put(2176,-4861){\makebox(0,0)[lb]{\smash{\SetFigFont{14}{16.8}{\rmdefault}{\mddefault}{\updefault}\special{ps: gsave 0 0 0 setrgbcolor}$R_1$\special{ps: grestore}}}}
\put(7876,-2761){\makebox(0,0)[lb]{\smash{\SetFigFont{14}{16.8}{\rmdefault}{\mddefault}{\updefault}\special{ps: gsave 0 0 0 setrgbcolor}$\alpha$\special{ps: grestore}}}}
\end{picture}

%% file: torus_23.pstex_t
\begin{picture}(0,0)%
\includegraphics{torus_23.pstex}%
\end{picture}%
\setlength{\unitlength}{1776sp}%
\begingroup\makeatletter\ifx\SetFigFont\undefined%
\gdef\SetFigFont#1#2#3#4#5{%
  \reset@font\fontsize{#1}{#2pt}%
  \fontfamily{#3}\fontseries{#4}\fontshape{#5}%
  \selectfont}%
\fi\endgroup%
\begin{picture}(12174,4842)(439,-5098)
\put(2176,-4861){\makebox(0,0)[lb]{\smash{\SetFigFont{14}{16.8}{\rmdefault}{\mddefault}{\updefault}\special{ps: gsave 0 0 0 setrgbcolor}$R$\special{ps: grestore}}}}
\put(8251,-661){\makebox(0,0)[lb]{\smash{\SetFigFont{14}{16.8}{\rmdefault}{\mddefault}{\updefault}\special{ps: gsave 0 0 0 setrgbcolor}{\bf B}\special{ps: grestore}}}}
\put(7501,-1936){\makebox(0,0)[lb]{\smash{\SetFigFont{14}{16.8}{\rmdefault}{\mddefault}{\updefault}\special{ps: gsave 0 0 0 setrgbcolor}$R$\special{ps: grestore}}}}
\put(7426,-4036){\makebox(0,0)[lb]{\smash{\SetFigFont{14}{16.8}{\rmdefault}{\mddefault}{\updefault}\special{ps: gsave 0 0 0 setrgbcolor}$R$\special{ps: grestore}}}}
\put(2401,-661){\makebox(0,0)[lb]{\smash{\SetFigFont{14}{16.8}{\rmdefault}{\mddefault}{\updefault}\special{ps: gsave 0 0 0 setrgbcolor}{\bf A}\special{ps: grestore}}}}
\put(601,-3161){\makebox(0,0)[lb]{\smash{\SetFigFont{14}{16.8}{\rmdefault}{\mddefault}{\updefault}\special{ps: gsave 0 0 0 setrgbcolor}$R$\special{ps: grestore}}}}
\end{picture}

%% file: angles.pstex_t
\begin{picture}(0,0)%
\includegraphics{angles.pstex}%
\end{picture}%
\setlength{\unitlength}{2368sp}%
\begingroup\makeatletter\ifx\SetFigFont\undefined%
\gdef\SetFigFont#1#2#3#4#5{%
  \reset@font\fontsize{#1}{#2pt}%
  \fontfamily{#3}\fontseries{#4}\fontshape{#5}%
  \selectfont}%
\fi\endgroup%
\begin{picture}(11413,5914)(75,-7685)
\put(751,-2086){\makebox(0,0)[lb]{\smash{\SetFigFont{14}{16.8}{\rmdefault}{\mddefault}{\updefault}\special{ps: gsave 0 0 0 setrgbcolor}$x^5$\special{ps: grestore}}}}
\put(10876,-5011){\makebox(0,0)[lb]{\smash{\SetFigFont{14}{16.8}{\rmdefault}{\mddefault}{\updefault}\special{ps: gsave 0 0 0 setrgbcolor}$N_a$\special{ps: grestore}}}}
\put(4501,-2536){\makebox(0,0)[lb]{\smash{\SetFigFont{14}{16.8}{\rmdefault}{\mddefault}{\updefault}\special{ps: gsave 0 0 0 setrgbcolor}$N_b$\special{ps: grestore}}}}
\put(11101,-7486){\makebox(0,0)[lb]{\smash{\SetFigFont{14}{16.8}{\rmdefault}{\mddefault}{\updefault}\special{ps: gsave 0 0 0 setrgbcolor}$x^4$\special{ps: grestore}}}}
\end{picture}

%% file: examples_12.pstex_t
\begin{picture}(0,0)%
\includegraphics{examples_12.pstex}%
\end{picture}%
\setlength{\unitlength}{2368sp}%
\begingroup\makeatletter\ifx\SetFigFont\undefined%
\gdef\SetFigFont#1#2#3#4#5{%
  \reset@font\fontsize{#1}{#2pt}%
  \fontfamily{#3}\fontseries{#4}\fontshape{#5}%
  \selectfont}%
\fi\endgroup%
\begin{picture}(8733,7614)(601,-6811)
\put(2851,-4861){\makebox(0,0)[lb]{\smash{\SetFigFont{10}{12.0}{\rmdefault}{\bfdefault}{\updefault}\special{ps: gsave 0 0 0 setrgbcolor}$1$\special{ps: grestore}}}}
\put(2926,-6661){\makebox(0,0)[lb]{\smash{\SetFigFont{10}{12.0}{\rmdefault}{\bfdefault}{\updefault}\special{ps: gsave 0 0 0 setrgbcolor}$2$\special{ps: grestore}}}}
\put(2926,-5911){\makebox(0,0)[lb]{\smash{\SetFigFont{10}{12.0}{\rmdefault}{\bfdefault}{\updefault}\special{ps: gsave 0 0 0 setrgbcolor}$3$\special{ps: grestore}}}}
\put(3001,-211){\makebox(0,0)[lb]{\smash{\SetFigFont{10}{12.0}{\rmdefault}{\bfdefault}{\updefault}\special{ps: gsave 0 0 0 setrgbcolor}$1$\special{ps: grestore}}}}
\put(2251,-211){\makebox(0,0)[lb]{\smash{\SetFigFont{10}{12.0}{\rmdefault}{\bfdefault}{\updefault}\special{ps: gsave 0 0 0 setrgbcolor}$2$\special{ps: grestore}}}}
\put(4651,-5986){\makebox(0,0)[lb]{\smash{\SetFigFont{10}{12.0}{\rmdefault}{\mddefault}{\updefault}\special{ps: gsave 0 0 0 setrgbcolor}$\Theta$\special{ps: grestore}}}}
\put(7726,-6811){\makebox(0,0)[lb]{\smash{\SetFigFont{10}{12.0}{\rmdefault}{\mddefault}{\updefault}\special{ps: gsave 0 0 0 setrgbcolor}$\Theta$\special{ps: grestore}}}}
\put(8476,-2011){\makebox(0,0)[lb]{\smash{\SetFigFont{10}{12.0}{\rmdefault}{\mddefault}{\updefault}\special{ps: gsave 0 0 0 setrgbcolor}$\Theta$\special{ps: grestore}}}}
\put(5476,-1336){\makebox(0,0)[lb]{\smash{\SetFigFont{10}{12.0}{\rmdefault}{\mddefault}{\updefault}\special{ps: gsave 0 0 0 setrgbcolor}$\Theta$\special{ps: grestore}}}}
\put(601,-61){\makebox(0,0)[lb]{\smash{\SetFigFont{11}{13.2}{\rmdefault}{\mddefault}{\updefault}\special{ps: gsave 0 0 0 setrgbcolor}$\mbox{\bf T}_1$\special{ps: grestore}}}}
\put(4051,-361){\makebox(0,0)[lb]{\smash{\SetFigFont{11}{13.2}{\rmdefault}{\mddefault}{\updefault}\special{ps: gsave 0 0 0 setrgbcolor}$\mbox{\bf T}_2$\special{ps: grestore}}}}
\put(6901,-361){\makebox(0,0)[lb]{\smash{\SetFigFont{11}{13.2}{\rmdefault}{\mddefault}{\updefault}\special{ps: gsave 0 0 0 setrgbcolor}$\mbox{\bf T}_3$\special{ps: grestore}}}}
\put(676,-4711){\makebox(0,0)[lb]{\smash{\SetFigFont{11}{13.2}{\rmdefault}{\mddefault}{\updefault}\special{ps: gsave 0 0 0 setrgbcolor}$\mbox{\bf T}_1$\special{ps: grestore}}}}
\put(3901,-4711){\makebox(0,0)[lb]{\smash{\SetFigFont{11}{13.2}{\rmdefault}{\mddefault}{\updefault}\special{ps: gsave 0 0 0 setrgbcolor}$\mbox{\bf T}_2$\special{ps: grestore}}}}
\put(6901,-4711){\makebox(0,0)[lb]{\smash{\SetFigFont{11}{13.2}{\rmdefault}{\mddefault}{\updefault}\special{ps: gsave 0 0 0 setrgbcolor}$\mbox{\bf T}_3$\special{ps: grestore}}}}
\end{picture}